\documentclass[prd,
 superscriptaddress, notitlepage,
 amsmath, amssymb,
 aps, preprintnumbers, nofootinbib, floatfix, longbibliography
]{revtex4-1}

\usepackage{graphicx}
\usepackage{dcolumn}
\usepackage{bm}
\usepackage{ulem}
\usepackage{color}
\usepackage{amssymb}
\usepackage{amsmath}
\usepackage{comment}
\usepackage{adjustbox}
\usepackage{rotating}
\usepackage{multirow}
\usepackage{graphicx}
\usepackage{lipsum}
\usepackage{hyperref}
 
\begin{document}
\preprint{}
\title{Detection prospects of long-lived quirk pairs at the LHC far detectors}

\author{Jinmian Li}
\email{jmli@scu.edu.cn}
\affiliation{College of Physics, Sichuan University, Chengdu 610065, China}

\author{Xufei Liao}
\email{xufeiliao12@outlook.com}
\affiliation{College of Physics, Sichuan University, Chengdu 610065, China}

\author{Jian Ni}
\email{jian.ni.2020@outlook.com}
\affiliation{College of Physics, Sichuan University, Chengdu 610065, China}

\author{Junle Pei}
\email{peijunle@ihep.ac.cn (corresponding author)}
\affiliation{Institute of High Energy Physics, Chinese Academy of Sciences, Beijing 100049, China}
\affiliation{Spallation Neutron Source Science Center, Dongguan 523803, China}



\begin{abstract}
We examine the sensitivity reaches of several LHC far detectors, such as FASER2, MATHUSLA, ANUBIS, SND@LHC, and FACET, to five simplified quirk scenarios. We include the next-to-leading order QCD corrections in our simulation of quirk events, which enhance the total production rate and increase the fraction of events in the forward direction for most cases. We calculate the time scales for the quirk pair to lose energy through radiations and for the quirk pair annihilation. Our results show that these far detectors can offer promising probes to the quirk scenario, complementing the searches at the main detectors. {Especially, FACET and FASER2 detectors can surpass the majority of searches conducted at the LHC main detector, with the exception of the HSCP search, for the color-neutral quirk $\mathcal{E}$.}
\end{abstract}

\maketitle



\section{Introduction}\label{sec:intro} 

Many models of new physics beyond the Standard Model (SM) introduce new SU(N) gauge symmetries along with the SM gauge group. These models, such as folded supersymmetry~\cite{Burdman:2006tz,Burdman:2008ek}, twin Higgs framework~\cite{Chacko:2005pe,Craig:2015pha,Serra:2019omd}, and quirky little Higgs~\cite{Cai:2008au}, can provide dark matter candidates, account for neutrino oscillation, and address the gauge hierarchy problem of the SM.
We call a particle a quirk if it is charged under both the SM gauge group and the new confining SU(N) gauge group and has a mass ($m_\mathcal{Q}$) much larger than the SU(N) confinement scale ($\Lambda$).
{Owing to the conservation of SU(N) gauge symmetry, quirks at colliders are exclusively produced in pairs as singlets of the new gauge group, facilitated by the Standard Model (SM) gauge couplings.} Unlike the QCD quark, the gauge flux tube between two quirks of a pair can have a macroscopic length~\cite{Kang:2008ea}, causing the quirk trajectory to oscillate during the propagation. The order of oscillation amplitude in the center of mass (c.m.) frame of the quirk pair  is roughly estimated as 
\begin{align}
\ell \sim \mathcal{O}(1)~\text{cm} \times (\frac{1~\text{keV}}{\Lambda})^{2} \times (\frac{m_\mathcal{Q}}{100~\text{GeV}})~.~ \label{eq:amp}
\end{align} 

Assuming the quirk mass around the electroweak scale, for the infracolor confinement scale $\Lambda \lesssim \mathcal{O}(10)$ eV, the infracolor force is too weak to induce any observable effects on the quirk trajectory due to the finite spatial resolution of the detector. The trajectory of each charged quirk traveling through the tracker system can only be reconstructed as a normal helical track. Such a signature is found to be stringently constrained by the heavy stable charged particle searches at the LHC~\cite{Farina:2017cts}. 
On the other hand, when $\Lambda \gtrsim \mathcal{O}(10)$ MeV, the quirk-pair system oscillates intensely after production and quickly loses its kinetic energy {through photon, (infracolor) glueball, and QCD hadron radiations}. {Upon transitioning to the ground state, the quirk pair may annihilate into SM particles, thereby imposing constraints derived from heavy resonance searches at the LHC~\cite{Cheung:2008ke,Harnik:2008ax,Harnik:2011mv,Fok:2011yc,Chacko:2015fbc,Capdevilla:2019zbx,Curtin:2021spx}.} 

The quirk signal becomes non-conventional for the $\Lambda \in [100~\text{eV}, 1~\text{MeV}]$. 
According to Eq.~\ref{eq:amp}, the quirk-pair system has a macroscopic oscillation amplitude for $\Lambda$ below keV. However, the string tension is large enough to prevent each quirk from following a helical trajectory in this $\Lambda$ range. The hits on trackers induced by the quirks are dropped in conventional event reconstruction at the LHC. Moreover, the heavy quirk deposits little energy in the electromagnetic/hadronic calorimeter~\cite{Li:2020aoq}. As a result, the missing quirk induces unbalanced transverse momentum in event reconstruction and is constrained by the mono-jet searches at the LHC~\cite{Farina:2017cts} if the quirk pair is produced with an energetic initial state radiated (ISR) jet recoiling against it. {However, when the momentum of the quirk-pair system is low, it can be stopped\footnote{This corresponds to $\lesssim \mathcal{O}(10)\%$ of total events for quirk with mass around $\mathcal{O}(100)$ GeV.} in the calorimeters due to ionization energy loss.} After many oscillations, the quirk pair annihilates at a time when there are no active $pp$ collisions and is constrained by the stopped long-lived particle (LLP) searches at the LHC~\cite{Evans:2018jmd,ATLAS:2013whh,CMS:2017kku}.
Some studies investigate the non-helical quirk trajectory further. Ref.~\cite{Knapen:2017kly} pointed out that the hits of quirk pair almost lie on a plane with a deviation smaller than $\mathcal{O}(100)~\mu$m if the quirk-pair system is moderately boosted and the infracolor force ($\Lambda \sim \mathcal{O}(1)$ keV) is much stronger than the Lorentz force ($B=4$ T in the CMS detector). Such a search based on coplanar hits has a high signal selection efficiency. Moreover, Ref.~\cite{Li:2019wce} found that the relatively large ionization energy loss of each quirk hit in the tracker can further improve the coplanar hits search.
For $\Lambda$ above $\mathcal{O}(10)$ keV, the quirk oscillation amplitude is no longer resolvable by detectors. The electrically neutral quirk-pair system leaves hits along a straight line inside the tracker, which is reconstructed as a single ultra-boosted charged particle with a high ionization energy loss. This signal was searched at the Tevatron~\cite{D0:2010kkd}, without giving evidence.

For $\Lambda < 1$ MeV, the quirk pair is long-lived and could be detected by far detectors around the LHC. 
Several operated/proposed experiments at the LHC search for LLPs~\cite{Anchordoqui:2021ghd,Knapen:2022afb,Garzelli:2022vad}.
The FASER detector~\cite{FASER:2018eoc,FASER:2022hcn} and SND@LHC~\cite{SHiP:2020sos,SNDLHC:2022ihg} detector are already installed and operated since the LHC Run-3. They are located in opposite forward directions with respect to the interaction point (IP). Both the detectors are sensitive to light long-lived BSM particles and neutrinos produced at the ATLAS IP and propagating for several hundred meters through the shields. 
The FACET detector~\cite{Cerci:2021nlb} follows a similar philosophy and will be located roughly 100 m forward from the CMS. It covers a larger pseudo-rapidity range than FASER and SND@LHC, though beam backgrounds may be more challenging to mitigate. 
The MATHUSLA~\cite{Curtin:2018mvb,MATHUSLA:2020uve} and ANUBIS~\cite{Bauer:2019vqk} detectors also aim to search for long-lived new physics particles at the LHC, but are located at large angles relative to the beamline, unlike the Forward Physics Facility (FPF) experiments. 
They are similar to the ATLAS and CMS muon chambers but have a much thicker shield. This results in much lower backgrounds. The typical sizes of the detectors are around $\mathcal{O}(100)$ m to ensure high geometric acceptance. 
{Due to the substantial mass of quirk particles, they do not experience significant showering within materials. When both quirks from a pair traverse through dense mediums like rock and concrete, the moving direction of the quirk-pair system remains nearly unchanged. This stability is attributed to the net electric charge of the quirk-pair system being zero, a conclusion supported by the findings in Ref.~\cite{Li:2021tsy}. The ionization energy loss for each quirk as it moves through a material is velocity-dependent, as detailed in Ref.~\cite{Li:2021tsy}. Consequently, a quirk particle in a pair produced at the ATLAS IP with high momentum will undergo minimal energy loss due to ionization while passing through such shields. An analysis focusing on the ratio of momentum loss in the direction of the quirk-pair motion through one meter of rock to the total momentum magnitude of the quirk-pair system has been conducted. It has been observed that approximately $50\%$ of events with quirk pair rapidity in the range of [-3,3] (corresponding to signal events at ANUBIS and MATHUSLA) are able to penetrate through 50 meters of rock without being stopped by ionization energy loss. Additionally, for forward events with a polar angle of the quirk pair smaller than 0.005, due to the significant momentum in the beam line direction, $70\%-80\%$ of these events are able to penetrate through 500 meters of rock.} At the hadron collider, the quirk pair is inevitably produced along with ISR jets. However, when the transverse momenta of the jets are small compared to the quirk mass, the quirk-pair system still flies in an almost forward direction. 
According to our study in Ref.~\cite{Li:2021tsy}, percent-level quirk events can reach the FASER2 detector which has a transverse radius of $1$ m. And the fraction is higher for larger $\Lambda$, as the oscillation amplitude becomes too small to bypass the detector. 
For more energetic initial state radiation (ISR), the quirk-pair system is pushed away from the forward direction. The MATHUSLA and ANUBIS detectors are more sensitive to detect events in this phase space. 
In this work, we will mainly focus on the parameter space with microscopic quirk oscillation amplitude, and consider the sensitivities of above-mentioned far detectors to the quirk particle. 
In such a case, two quirks from one pair can reach the detectors simultaneously. The additional information from the energetic track pair and relatively smaller velocity compared to the muon renders the quirk signal to be almost background free. {In TABLE~\ref{tab:quirkconstrains}, we provide a summary of the primary signatures and search methodologies across various confinement scales.}

This paper is organized as follows. In Sec.~\ref{sec:model}, we introduce five simplified quirk model frameworks and discuss the cross-sections and kinematics for the production at the LHC. Section~\ref{sec:evo} provides the calculation details for each stage of quirk pair evolution after the production. The setup for the detectors in our simulation is explained in Sec.~\ref{sec:det}. The sensitivity reaches of all those far detectors for five quirk scenarios are presented in Sec.~\ref{sec:exc}. Finally, we draw the conclusion in Sec.~\ref{sec:conclude}.

\begin{table}[tbp]
	\centering 
	\begin{tabular}{|c|c|c|c|} \hline
	confinement scale  & signature & search method  \\ \hline
	$\lesssim \mathcal{O}(10)$ eV & track with high ionization energy loss & HSCP search~\cite{Farina:2017cts}   \\ \hline
	$[100~\text{eV}, 1~\text{MeV}]$ & anomalous tracks recoiling energetic ISR & mono-jet search~\cite{Farina:2017cts} \\  \hline
	$[100~\text{eV}, 1~\text{MeV}]$ &  out-of-time annihilation & stopped LLP search~\cite{Evans:2018jmd} \\  \hline
	$\sim \mathcal{O}(1)$ keV & non-helical coplanar tracks & coplanar hits (large ionization) search~\cite{Knapen:2017kly}\\ \hline
	$[10~\text{keV}, 1~\text{MeV}]$ & straight highly-ionizing track &  single highly ionizing track search~\cite{D0:2010kkd} \\ \hline
	$[10~\text{keV}, 1~\text{MeV}]$ & long-lived charged particles & searches at the LHC far detectors$[$This work$]$   \\ \hline
	$\gtrsim \mathcal{O}(10)$ MeV & prompt decay into SM particles & heavy resonances search~\cite{Harnik:2011mv,Capdevilla:2019zbx} \\ \hline
	\end{tabular}
	\caption{\label{tab:quirkconstrains} Bounds on quirk from different searches.  }
\end{table}

\section{Simplified quirk models}\label{sec:model}

We consider two classes of models. In the first class, a single quirk is introduced in addition to the SM particle content. 
The quirk can be either a fermion or a scalar with the same SM quantum numbers as a right-handed charged lepton or a right-handed down-type quark. 
There will be four scenarios depending on the quirk quantum number, {\it i.e.} under $SU(N_{\text{IC}}) \times SU_C(3) \times SU_L(2) \times U_Y(1)$ gauge group: 
\begin{align}
	\text{Scalar:}~\tilde{\mathcal{D}} &= \left( N_{\text{IC}}, 3, 1, -1/3 \right),\\
	\text{Scalar:}~\tilde{\mathcal{E}} &= \left( N_{\text{IC}}, 1, 1, -1 \right),\\
	\text{Fermion:}~\mathcal{D} &= \left( N_{\text{IC}}, 3, 1, -1/3 \right),\\
	\text{Fermion:}~\mathcal{E} &= \left( N_{\text{IC}}, 1, 1, -1 \right) \label{eq::quannum}.
\end{align}
In the second class, we extend the SM with a right-handed $SU_R(2)$ gauge group, where the gauge boson $W_R$ couples to quirks (denoted by the $W_R$ scenario hereafter).
This scenario illustrates the case where quirk annihilation is slow when $W_R$ is relatively heavy. {Since we only consider the production of quirk pairs with different flavors in this scenario~\footnote{The production of quirk pairs with the same flavor is similar to those scenarios in the first class. The annihilation of quirk pair of the same flavor is always prompt.}, the colored quirk has similar properties as the color-neutral one except for the color factors in production and annihilation}. So we study the color-neutral scalar quirk, whose annihilations to the SM fermions are $p$-wave suppressed. 
The relevant Lagrangian is given as 
\begin{align}
\mathcal{L} = \kappa g_2 \cdot W^+_{R \mu}  \left( \bar{u}_i \gamma^\mu P_R d_i - i (\partial^\mu \bar{\tilde{\mathcal{V}}}) \tilde{\mathcal{E}} + i (\partial^\mu \tilde{\mathcal{E}} ) \bar{\tilde{\mathcal{V}}}   \right)+\text{H.c.},
\end{align}
where $u_i$ and $d_i$ are SM quarks and $i$ is the generation index. The $\tilde{\mathcal{V}}$ and $\tilde{\mathcal{E}}$ are scalar quirk fields that carry the same SM charges as right-handed neutrino (with hypercharge $Y=0$) and lepton, respectively. $g_2$ is the SM weak coupling constant and $\kappa$ is an normalization factor. 

We choose $N_{\text{IC}}=2$ for the infracolor gauge group in this work for simplicity. 
Note that quirk pair production cross sections are proportional to $N_{\text{IC}}$. Moreover, the rates of quirk annihilation and infracolor glueball radiation also depend on $N_{\text{IC}}$, as we will show later. 
We implement the Monte Carlo simulation of quirk events at the LHC in the \texttt{MG5\_aMC@NLO} framework~\cite{Alwall:2014hca}, which uses model files of the five scenarios mentioned above written by \texttt{FeynRules}~\cite{Alloul:2013bka}. 
The \texttt{Pythia8}~\cite{Sjostrand:2007gs} is used for simulating the parton shower of both SM particles and colored quirks. 
The hadronization is turned off, but we expect that it will not significantly change the momenta of the quirks as their masses are much larger than those of quarks.
On the other hand, only the quirk-quark bound states are observable particles at detectors due to the color confinement. It was found that for the $\tilde{\mathcal{D}}^c$ and $\mathcal{D}^c$ quirks, the probability of the quirk-quark bound state pair to have charge $\pm 1$ is around 30\%~\cite{Knapen:2017kly}. 
Since only electric-charged quirk final states can induce visible signals at far detectors, we include this hadronization effect by multiplying the production rate by a factor of $1/3$ for $\tilde{\mathcal{D}}^c$ and $\mathcal{D}^c$.

\subsection{Production at the LHC}
\begin{figure}[t!]
\includegraphics[width=0.32\textwidth]{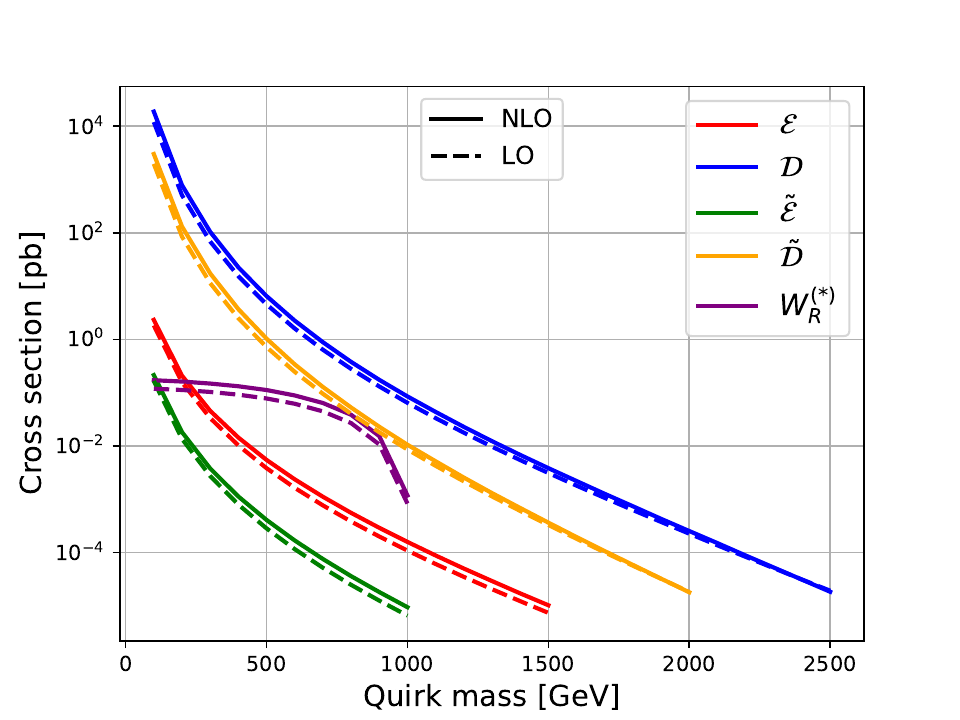}
\includegraphics[width=0.32\textwidth]{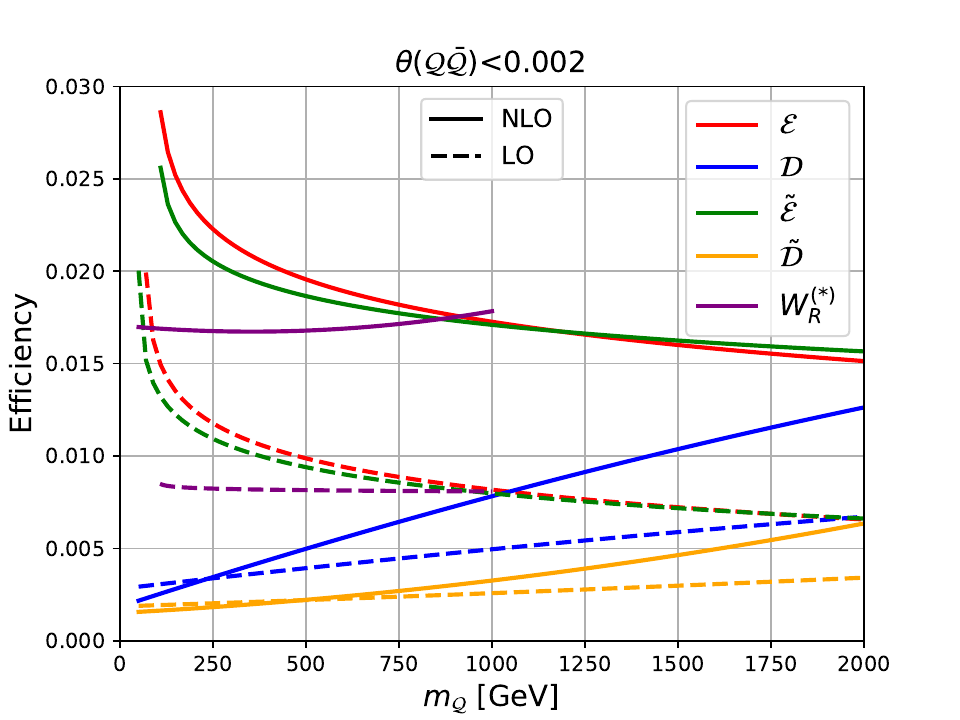}
\includegraphics[width=0.32\textwidth]{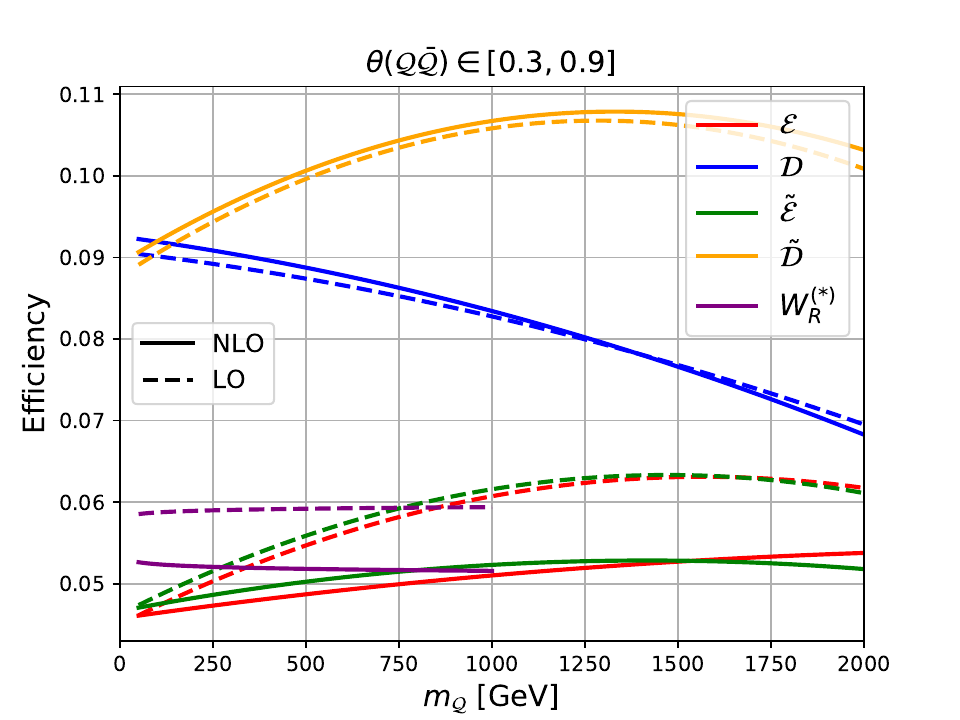}
\caption{Left panel: the LO and NLO total production cross sections for different quirk scenarios. We take $\kappa=1$ and $m_{W_R}=2$ TeV for the $W_R$ scenario. 
The middle and right panels show the fraction of events that have $\theta(\mathcal{Q} \bar{\mathcal{Q}})<0.002$ and $\theta(\mathcal{Q} \bar{\mathcal{Q}}) \in [0.3,0.9]$ for varying quirk mass in all scenarios. \label{fig.xsec}}
\end{figure}

Quirk pairs can be produced at the LHC through their SM gauge interactions, as well as the $SU_R(2)$ gauge interaction in the $W_R$ scenario. 
The QCD processes with gluon exchange dominate the production of colored quirks, while Drell-Yan processes with SM $\gamma/Z$ or $W_R$ exchange produce color-neutral quirks.
We use the NLO QCD function of \texttt{MG5\_aMC@NLO} to calculate the production cross-sections for the $\tilde{\mathcal{D}}$, $\tilde{\mathcal{E}}$, $\mathcal{D}$, $\mathcal{E}$ and $W_R$ scenarios. 
The left panel of Fig.~\ref{fig.xsec} shows the LO and NLO QCD cross sections for comparison, where we take $\kappa=1$ and $m_{W_R}=2000$ GeV for the $W_R$ scenario.
The quirk mass exceeding $m_{W_R}/2$ in the $W_R$ scenario dramatically reduces the production cross section as the $W_R$ becomes off-shell. 
The NLO QCD effects can increase the production rate by a factor of $1\sim 2$ depending on the quirk mass for all scenarios.

The polar angle of the quirk-pair system ($\theta(\mathcal{Q} \bar{\mathcal{Q}})$) is the most relevant kinematic variable that determines the detection efficiencies. 
For example, the FASER2 detector is designed with a radius of 1 m and is located 480 m downstream from the ATLAS IP aligned with the beam collision axis. So, only events with $\theta(\mathcal{Q} \bar{\mathcal{Q}} \lesssim 1/480 \sim 0.002)$ can potentially trigger signals at FASER2. 
In the middle and right panels of Fig.~\ref{fig.xsec}, we plot the fractions of events that have $\theta(\mathcal{Q} \bar{\mathcal{Q}})<0.002$ (corresponding to FASER2) and $\theta(\mathcal{Q} \bar{\mathcal{Q}}) \in [0.3,0.9]$ (corresponding to MATHUSLA) for different quirk masses in various scenarios. 
The non-zero $\theta(\mathcal{Q} \bar{\mathcal{Q}})$ is attributed to ISR and final state radiation (FSR) in quirk production processes. 
The signal efficiencies of colored quirks are lower than those of color-neutral quirks at the FASER detector because they undergo more intense FSRs that deflect the direction of the quirk pair. The opposite is true for the MATHUSLA detector. 
However, this FSR effect deflects the momentum direction less for heavier quirks.
So, there will be more signal events flying along the forward direction for heavier colored quirk when the FSR dominates over the ISR. 
Only the ISR is relevant for color-neutral quirks, and its energy scale is proportional to the quirk masses. 
Heavier quirk can produce more energetic ISR, leading to more events with larger $\theta(\mathcal{Q} \bar{\mathcal{Q}})$. 
As a result, the signal efficiencies depend on quirk mass oppositely for colored and color-neutral quirks.
For the $W_R$ scenario, the quirk pair production is dominated by the on-shell $W_R$ decay. So, the dependence on quirk mass is negligible.  
The NLO QCD correction tends to increase the quirk production rates in the forward direction for all scenarios. And the increase is more significant for color-neutral quirks. 
A similar conclusion has also been drawn in Refs.~\cite{Cullen:2012eh,Backovic:2015soa,Ruiz:2015zca,Fuks:2016ftf}, which study the production of other BSM particles with the same quantum number. 

\begin{figure}[htbp]
\includegraphics[width=0.48\textwidth]{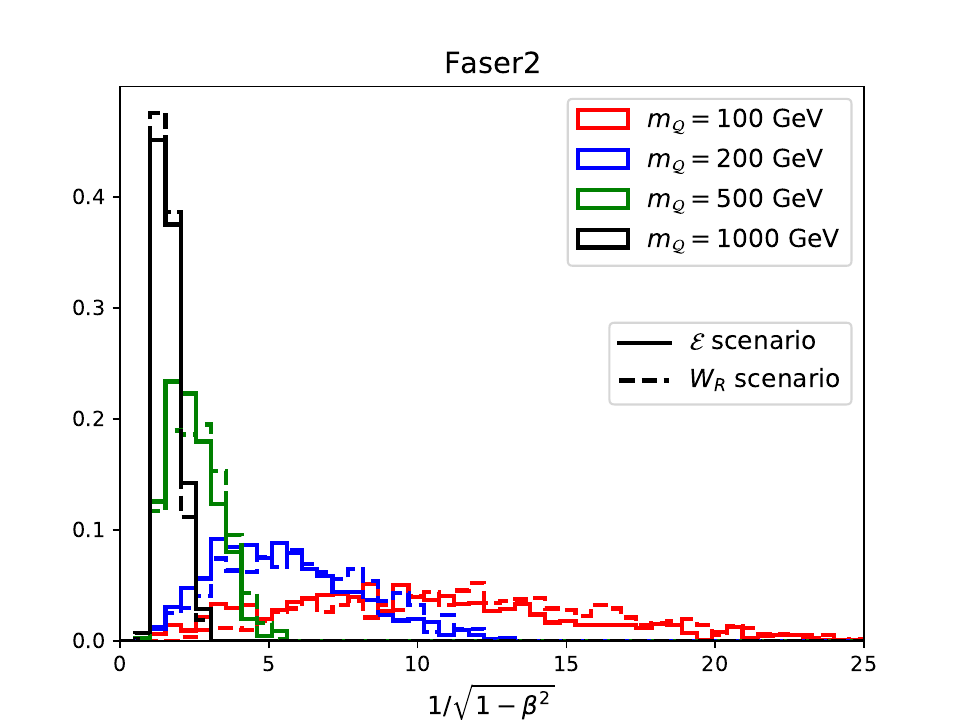}
\includegraphics[width=0.48\textwidth]{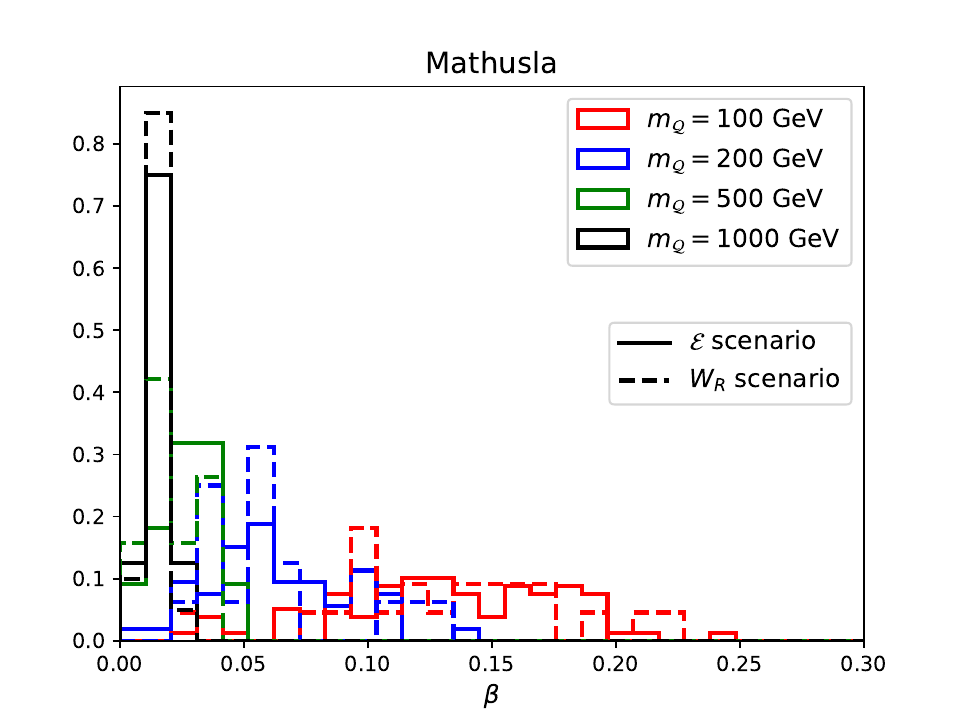}
\caption{Normalized distributions of the boost factor for events in FASER2 detector (left) and the distribution of the velocity for events in the MATHUSLA detector (right).  For the $W_R$ scenario, we have chosen $m_{W_R} =2000$ GeV. \label{fig.gamma}}
\end{figure}

The traveling distance of the quirk pair in the laboratory frame also depends on the velocity ($\beta$) / boost factor {($\gamma=1/\sqrt{1-\beta^2}$)} of the quirk-pair system. In Fig.~\ref{fig.gamma}, we present the distribution of the boost factor for events in FASER2 detector and the distribution of the velocity for events in the MATHUSLA detector~\footnote{We present velocities for MATHUSLA case because the boost factors here are too close to unit. }. 
For the quirk-pair system, the longitudinal momentum can be easily induced by the imbalance between the momenta of two initial partons, leading to relatively high boost factors of quirk events at the FASER2 detector. 
In contrast, the transverse momentum of the quirk-pair system can only be obtained through initial/final state radiations. The rate of energetic radiation is suppressed in the quirk production process. Given the quirk mass of $\mathcal{O}(100)$ GeV, most of the events at MATHUSLA only have small velocities $\beta \lesssim 0.2$. 
Distributions for quirks with different masses are shown in different colors. It is obvious that the velocity/boost factor is decreased with increasing quirk mass. For $m_{\mathcal{Q}} \gtrsim 500$ GeV, the boost factor becomes $\gamma \lesssim 3$ for most of the events. 
The distributions are quite similar among $\tilde{\mathcal{D}}$, $\tilde{\mathcal{E}}$, $\mathcal{D}$ and $\mathcal{E}$ scenarios. So, only those in the $\mathcal{E}$ scenario is presented. The quirk in $W_R$ scenario is dominantly produced from $W_R$ decay, the kinematics of which is different from other scenarios. As a result, the boost factor in the $W_R$ is slightly enhanced.

\section{The lifetime of quirk pair} \label{sec:evo}

At the LHC, the quirk is produced with some kinetic energy. {The quirk energy loss can proceed through the radiations of infra-color glueballs,  photons~\cite{Evans:2018jmd}, and QCD hadrons.} 
The quirk pair will eventually settle into a quirkonium ground state which decays through constituent quirk annihilation. 
Moreover, there are two quirks in the $W_R$ scenario. The heavier one can decay into the lighter one through $W_R$ with decay width related to the $W_R$ mass and the mass splitting between two quirks. 
In this section, we discuss each of those processes. 

\subsection{Energy loss through radiation}

\subsubsection{Infracolor glueball emission}
We follow the assumption as adopted in Refs.~\cite{Evans:2018jmd,Harnik:2011mv} to describe the infracolor glueball radiation, that is during each oscillation, the quirk pair has a probability $\epsilon$ to emit an infracolor glueball with the energy of $\Lambda$, leading to
\begin{align}
\frac{dE}{dt} = -\frac{\epsilon \Lambda}{T}~, \label{eq:eloss1}
\end{align}
where $T$ is the quirk oscillating period.  
We will focus on a relatively large confinement scale in this work, {\it i.e.}, $\Lambda \gtrsim \mathcal{O}(1)~\text{keV}~(\sim \text{\AA}^{-1})$.
Just after the quirk is produced, the kinetic energy of quirk pair in the  center of mass (c.m.) frame $E_0$ is around $\mathcal{O}(100)$ GeV scale, leading to quirk oscillation amplitude $r \gg \Lambda^{-1}$. 
The interaction between two quirks is described by a linear potential $V(r) \sim \Lambda^2 r$. So, the oscillation period can be simply estimated by $T=2\Lambda^{-2}\sqrt{2m_\mathcal{Q} E_0}$. 
Using Eq.~(\ref{eq:eloss1}), we can obtain the time required to lose the energy of $\Delta E$ is
\begin{align}
\tau^{\text{IC}}_{\text{linear}} \sim  \int_{E_0}^{\sqrt{m_\mathcal{Q} \Lambda}}\frac{dE}{dE/dt}=\frac{4 \sqrt{2 m_\mathcal{Q}}(E_0^{3 / 2}-\sqrt{m_\mathcal{Q} \Lambda}^{3/2})}{3 \epsilon \Lambda^{3}}~. \label{eq:tlic}
\end{align}

The amplitude of quirk oscillation keeps shrinking due to the radiation. When the amplitude $r \lesssim \Lambda^{-1}$ (correspond to kinetic energy $\sim \sqrt{m_\mathcal{Q} \Lambda}$), the quirk potential becomes Coulombic, $V(r)\sim -\alpha_{\text{IC}}(r)/r$, where the running gauge coupling $\alpha_{\text{IC}}(r)=\frac{1}{b_{0}\log(1/(r^{2}\Lambda^{2}))}$. 
Taking the flavor number $N_f=1$ for $\mathcal{D}$, $\mathcal{E}$ scenarios, $N_s=1$ for $\tilde{\mathcal{D}}$, $\tilde{\mathcal{E}}$ scenarios, and $N_s=2$ for $W_R$ scenario, the $b_{0}=\frac{1}{4\pi}(11/3 N_{\text{IC}}-2/3 N_f-1/6 N_s)$. 
Similar as Eq.~(\ref{eq:eloss1}), with the oscillation period dictated by Kepler's law $T=\pi\sqrt{m_\mathcal{Q} r^3 \alpha_{\text{IC}}^{-1}}$, the change of the quirk binding energy $B=-V(r)$ can be expressed as 
\begin{align}
\frac{dB}{dt}=\frac{\epsilon\Lambda B^{3/2}}{\pi \alpha_{\text{IC}} \sqrt{m_\mathcal{Q}}}~.~
\end{align}
The time required to evolve the binging energy from initial $B_i \sim \Lambda$ to final $B_f=\alpha^2_{\text{IC}} m_\mathcal{Q}$ can is calculated as 
\begin{align}
\tau^{\rm IC}_{\rm Coulomb}=\int_{B_i}^{B_f} \frac{dB}{dB/dt} =\frac{2\pi \alpha_{\text{IC}} \sqrt{m_\mathcal{Q}}}{\epsilon \Lambda^{3/2}} - \frac{2 \pi}{\epsilon \Lambda}~.~
\end{align}

\subsubsection{Electromagnetic radiation}
The oscillating charged quirk can also lose energy through electromagnetic radiation. 
{As discussed in Ref.~\cite{Evans:2018jmd}}, for confinement scale $\Lambda	\gtrsim \text{\AA}^{-1}$, the energy loss via the photon radiation can be described by the Larmor formula 
\begin{align}
\frac{dE_{\gamma}}{dt} = \frac{16\pi Q^2}{3} \alpha_{\text{EM}} a^2~,
\end{align}
where $Q$ and $a=-\frac{d V(r)}{d r} \frac{1}{m_{\mathcal{Q}}}$ are the electric charge and the magnitude of the acceleration of the quirk.
Similar to the infracolor glue ball radiation, the quirk potential begins in a linear regime as long as the quirk oscillation amplitude $r \gtrsim \Lambda^{-1}$. The time required to lose energy of $\Delta E$ until $r \sim \Lambda^{-1}$ is
\begin{align}
\tau^{\rm EM}_{\rm linear}=\int_{E_0}^{\sqrt{m_\mathcal{Q} \Lambda}}\frac{dE}{dE/dt}=\frac{3 (E_{0}-\sqrt{m_\mathcal{Q} \Lambda}) m_\mathcal{Q}^2}{16 Q^2 \pi \alpha_{\text{EM}} \Lambda^4}~.
\end{align}

In the Coulombic regime, where $r \lesssim \Lambda^{-1}$, the change of the quirk binding energy $B=-V(r)$ is given by
{
\begin{align}
\frac{dB}{dt} \sim \frac{16\pi Q^2 \alpha_{\text{EM}} B^4}{3\alpha^2_{\text{IC}} m^2_\mathcal{Q}}~.
\end{align}}
The time required to evolve the binding energy from initial $B_i \sim \Lambda$ to final $B_f=\alpha^2_{\text{IC}} m_\mathcal{Q}$ can be calculated as
{
\begin{align}
\tau^{\rm EM}_{\rm Coulomb}=\int_{B_i}^{B_f}\frac{dB}{dB/dt}=\frac{\alpha_{\text{IC}}^2 m_\mathcal{Q}^2}{16 Q^2 \pi \alpha_{\text{EM}} \Lambda^3}-\frac{1}{16 Q^2\pi \alpha_{\text{EM}} \alpha_{\text{IC}}^4 m_\mathcal{Q}}~.
\end{align}}

Considering the effects of both infracolor glueball radiation and the electromagnetic radiation, the overall time for quirk to lose its kinetic energy (in the CoM frame) is given by
\begin{align}
\tau_{\rm rad}=((\tau^{\rm IC}_{\rm linear})^{-1}+(\tau^{\rm EM}_{\rm linear})^{-1})^{-1}+((\tau^{\rm IC}_{\rm Coulomb})^{-1}+(\tau^{\rm EM}_{\rm Coulomb})^{-1})^{-1}~. 
\end{align}
Boosting to the laboratory (Lab) frame, we obtain
\begin{align}
t^{\rm lab}_{\rm rad}=\frac{\tau_{\rm loss}}{\sqrt{1-v_{\bar{\mathcal{Q}} \mathcal{Q}}^2/c^2}}~, \label{tlabrad}
\end{align}
where $v_{\bar{\mathcal{Q}} \mathcal{Q}}$ is the velocity of quirk-pair system in the laboratory frame. 
{
\subsubsection{QCD radiation}
When considering the colored quirks, $\tilde{\mathcal{D}}$ and $\mathcal{D}$, we assume that for each oscillation, the quirk pair possesses a probability $\epsilon^\prime$ of emitting hadrons and gluons, each with an energy of $E_\text{QCD}\sim$ 0.1 GeV. So, we get
\begin{align}
	\frac{dE}{dt} = -\frac{E_\text{QCD} \epsilon^\prime}{T}~. \label{eq:eloss2}
\end{align}
Similarly, we obtain
\begin{align}
	\tau^{\text{QCD}}_{\text{linear}} \sim  \int_{E_0}^{\sqrt{m_\mathcal{Q} \Lambda}}\frac{dE}{dE/dt}=\frac{4 \sqrt{2 m_\mathcal{Q}}(E_0^{3 / 2}-\sqrt{m_\mathcal{Q} \Lambda}^{3/2})}{3 \epsilon^\prime E_\text{QCD} \Lambda^{2}} \label{eq:tlic2}
\end{align}
for the linear quirk potential
and
\begin{align}
	\tau^{\rm QCD}_{\rm Coulomb}=\int_{B_i}^{B_f} \frac{dB}{dB/dt} =\frac{2\pi \alpha_{\text{IC}} \sqrt{m_\mathcal{Q}}}{\epsilon^\prime E_\text{QCD} \Lambda^{1/2}} - \frac{2 \pi}{ \epsilon^\prime E_\text{QCD}}
\end{align}
for the Coulombic quirk potential, respectively.
Upon incorporating the effects of QCD radiation, the total time required for energy dissipation in the CoM frame becomes 
\begin{align}
	\tau_{\rm tot}=((\tau^{\rm IC}_{\rm linear})^{-1}+(\tau^{\rm EM}_{\rm linear})^{-1}+(\tau^{\rm QCD}_{\rm linear})^{-1})^{-1}+((\tau^{\rm IC}_{\rm Coulomb})^{-1}+(\tau^{\rm EM}_{\rm Coulomb})^{-1}+(\tau^{\rm QCD}_{\rm Coulomb})^{-1})^{-1}~. \label{final}
\end{align}
}

{
\subsubsection{The probabilities of infracolor glueball emission and QCD radiation}\label{ppp}
According to \cite{Kang:2008ea}, the emission of infracolor glueballs and QCD radiation by the quirk pair is contingent upon the minimum separation distance between the two quirks, denoted as $d_\text{min}$, being less than $1/\Lambda_\text{QCD}$ and $1/\Lambda$, respectively. As the quirk pair oscillates, both the Lorentz force acting on the quirks in a magnetic field and the force resulting from ionization energy loss as a charged quirk moves through materials can impart a non-zero total angular momentum to the quirk pair, thereby ensuring that $d_\text{min}>0$. In this context, we use the minimum distance between two quirks after one oscillation in the
magnetic field ($\vec{B}$) to estimate $d_\text{min}$, which is approximately given by
  \begin{align}
    d_{\text{min}}=&1.2\times 0.231\times 10^{11}[\mu\text{m}]\nonumber\\
    & \times{\frac{m_\mathcal{Q}}{[\text{GeV}]} \left(\frac{\Lambda}{[\text{eV}]}\right)^{-4}\left(\sqrt{1+\rho^2}-\frac{\sinh^{-1}{\rho}}{\rho}\right)} {\frac{\left|q\left(\vec{\beta}\times \frac{\vec{B}}{[\text{T}]}\right)\cdot\hat{e}_3\right|}{\sqrt{1-\beta^2}}}{\sqrt{1+\frac{\beta^2}{1-\beta^2}\cos^2\alpha}}~,\label{dmin}
  \end{align}
where 
  \begin{align}
      &\vec{\beta}=(\vec{p}_1+\vec{p}_2)/(E_1+E_2)~,~~~~~~~\beta=|\vec{\beta}|~,\\
      & \rho=\sqrt{\frac{(E_1+E_2)^2-|\vec{p}_1+\vec{p}_2|^2}{4 m_\mathcal{Q}^2}-1}~,\\
      & \cos\alpha= \hat{e}_3\cdot \vec{\beta}/\beta~.\label{alpha}
  \end{align}
  $\vec{p}_i$ and $E_i$ ($i=1,2$) represent the initial three-momentum and energy of the quirks immediately after their production, respectively. $q$ denotes the electric charge of a quirk. The unit vectors $\hat{e}_1$, $\hat{e}_2$, and $\hat{e}_3$ have the same directions of $\vec{\beta}\times \vec{B}$, $(\vec{p}_1/E_1-\vec{p}_2/E_2)$, and $(\hat{e}_1-\hat{e}_1\cdot\hat{e}_2~ \hat{e}_2)$, respectively. A  derivation of $d_{\text{min}}$ is provided in Appendix~\ref{dp}.
  
Employing the formula in Eq.~(\ref{dmin}) to estimate the actual $d_\text{min}$ is a cautious approach. First, this estimation accounts for the Lorentz force acting on quirks due to the magnetic field. However, it does not factor in the forces resulting from ionization energy loss experienced by a charged quirk as it moves through materials, despite the fact that long-lived quirks predominantly traverse such environments. Second, as per \cite{Li:2020aoq}, the total angular momentum of the quirk pair is expected to increase significantly as they enter or exit the magnetic field or materials. This is attributed to scenarios where only one quirk of the pair is within the magnetic field or material at any given time. In our calculations, we simplify this by presuming that both quirks reside within the magnetic field for the duration of a single oscillation. Third, the emission of the infracolor glueball can also alter the momentum of the quirk pair due to the spin carried by the infracolor glueball.

In this study, we compare the minimum distance between two quirks after one oscillation in the magnetic field, denoted as $d^s_{\text{min}}$, obtained through simulating quirk motions using the method proposed in \cite{Li:2019wce}, with the distance calculated using Eq.~(\ref{dmin}), represented as $d_\text{min}$. We observe that approximately $88\%$ of quirk-pair events exhibit a ratio of $d_\text{min}/d^s_\text{min}$ within the range of [0.5, 1.5]. Consequently, we utilize $d_\text{min}$ as calculated by Eq.~(\ref{dmin}) to explore the parameter space of $\Lambda$ where the emission of infracolor glueballs and QCD radiation by the quirk pair is feasible. Our findings indicate that for the majority of events, achieving $d_\text{min}<1/\Lambda$ necessitates $\Lambda$ values exceeding several keV, whereas $d_\text{min}<1/\Lambda_\text{QCD}$ requires $\Lambda$ to be above approximately 0.1 MeV. In this work, we consider $\Lambda$ values ranging from keV to MeV. We adopt an $\epsilon$ value of approximately 0.1 for infracolor glueball emissions from quirk pairs, as suggested in \cite{Evans:2018jmd}, since most events can exhibit infracolor glueball emissions with $\Lambda$ values spanning from keV to MeV. Considering that QCD radiation from quirk pairs can manifest for $\Lambda$ values greater than 0.1 MeV, this study simplifies the analysis by either disregarding QCD radiation (setting $\epsilon^\prime$ to 0) or adopting a reduced value of approximately 0.01 for $\epsilon^\prime$ to account for QCD radiations from quirk pairs in the keV to MeV $\Lambda$ range.
}

\subsection{Quirk annihilation}

The quirk pair settles into a ground state through the radiation of infra-color glueballs and photons. 
The decay of the bound state ($B$) can be calculated in a similar way as meson decay in QCD~\cite{Barger:1987xg}. The partial decay width can be expressed as 
\begin{align}
\Gamma(B \to X) = \sigma v (\mathcal{Q} \bar{\mathcal{Q}} \to X) \times \left|\psi(0)\right|^2~,  
\end{align}
where $X$ is annihilation final states including SM particles and the infracolor gluon, $\sigma v (\mathcal{Q} \bar{\mathcal{Q}} \to X)$ is the cross-section of s-wave quirk annihilation in the CoM frame of $\mathcal{Q} \bar{\mathcal{Q}}$, and $\psi(0)$ is the wave function of quirkonium bound state evaluated at origin. 

A charged quirk pair made up of two quirks of the same flavor can annihilate into a pair of infracolor gluon and the photon. The fermionic quirk pair can also annihilate into the SM fermions. 
As for quirk carrying QCD color, it can annihilate into a pair of QCD gluon or $g \gamma$. The cross sections of those channels for the s-wave quirk annihilation in the CoM frame are provided in Tab.~\ref{tab:anni}. 
The parameters $\alpha_{\text{EM}}$, $\alpha_S$, and $\alpha_{\text{IC}}$ represent the electromagnetic coupling, the strong coupling, and the infracolor coupling, respectively. Note that the value of $\alpha_{\text{IC}}$ at the scale of $2 m_{\mathcal{Q}}$ will be used in our numerical study. The $N_{\text{IC}}=2$ and $N_C=3$ are considered. 

The ground state wave function $\psi(0)$ can be obtained by solving the stationary Schrodinger equation with a potential function $V(r)=- \alpha_{\text{IC}}(r) / r$.  
Since the potential is spherically symmetric, we only need to solve the radial differential equation with the separation of variables method: 
\begin{align}
y^{\prime\prime}_{n,l}(r)=[V_{eff}(r)-\varepsilon_{n,l}] y_{n,l}(r)~, \label{eq:sch}
\end{align}
where $y_{n,l}(r) = r R_{n,l}(r)$ with $R_{n,l}(r)$ is the radial wave function. It satisfies the normalization condition $\int_{0}^{\infty} dr [y_{n,l}(r)]^2 = 1$. 
The effective potential $V_{eff}(r) = 2\mu V(r) + \frac{l(l+1)}{r^2}$ includes the reduced mass $\mu$ and the angular momentum quantum number $l$. 
The corresponding eigenvalue is defined as $\varepsilon_{n,l} \equiv 2\mu E_{n,l}$, where $E_{n,l}$ is the energy eigenvalue. For ground state, $n=0$ and $l=0$. 
In our numerical calculation, the value of $\varepsilon_{n,l}$ is increased from 0 with a small step size. 
For each chosen $\varepsilon_{n,l}$, the Eq.~(\ref{eq:sch}) can be solved numerically by using the initial condition around the origin $y(\delta) = \delta^{l+1}$ and $y^\prime(\delta) = (l+1) \delta^l$, for $\delta \ll 1$. The correct eigenvalue of $\varepsilon_{n,l}$ can be identified when the asymptotic boundary condition $\lim\limits_{r \to \infty} y(r) \to 0$ is fulfilled. 

\begin{table}[tbp]
	\centering 
	\begin{tabular}{|c|c|c|} \hline
		Annihilation Process &  Fermion quirk & Scalar quirk \\ \hline
		$\mathcal{Q} \bar{\mathcal{Q}} \to g^\prime g^\prime$ &  $\frac{\pi\alpha^2_{\rm IC}(N_{\rm IC}^2-1)}{4m^2 N_C N_{\rm IC}}$ & $\frac{\pi\alpha^2_{\rm IC}(N_{\rm IC}^2-1)}{2m^2 N_C N_{\rm IC}}$ \\ \hline
		$\mathcal{Q} \bar{\mathcal{Q}} \to \gamma \gamma$ &  $\frac{N_{\rm IC}\pi \alpha_{\rm EM}^2 Q^4}{N_C m^2}$ & $\frac{2 N_{\rm IC}\pi \alpha_{\rm EM}^2 Q^4}{N_C m^2}$ \\ \hline
		$\mathcal{Q} \bar{\mathcal{Q}} \to \gamma \to ee$ &  $\frac{N_{\rm IC}\pi \alpha_{\rm EM}^2 Q^2}{N_C m^2}$ & $-$ \\ \hline
		$\mathcal{Q} \bar{\mathcal{Q}} \to uu$ &  $\frac{\pi\alpha^2_{s}(N_{C}^2-1) N_{\rm IC}}{4m^2 N_C }$ & $-$ \\ \hline
		$\mathcal{Q} \bar{\mathcal{Q}} \to g g$ & $\frac{N_{\rm IC}\pi\alpha_S^2 (N_C^4-3N^2+2)}{8m^2 N_C^3}$ & $\frac{N_{\rm IC}\pi\alpha_S^2 (N_C^4-3N^2+2)}{4m^2 N_C^3}$ \\ \hline
		$\mathcal{Q} \bar{\mathcal{Q}} \to g \gamma$ & $\frac{N_{\rm IC}\pi\alpha_{\rm EM}\alpha_S (N_C^2-1)Q^2}{m^2 N_C^2}$ & $ \frac{2 N_{\rm IC}\pi\alpha_{\rm EM}\alpha_S (N_C^2-1)Q^2}{m^2 N_C^2}$\\ \hline
	\end{tabular}
	\caption{\label{tab:anni} Cross-section of s-wave quirk annihilation in the center of mass frame of $\mathcal{Q} \bar{\mathcal{Q}}$ system. }
\end{table}

\begin{figure}[htbp]
\includegraphics[width=0.48\textwidth]{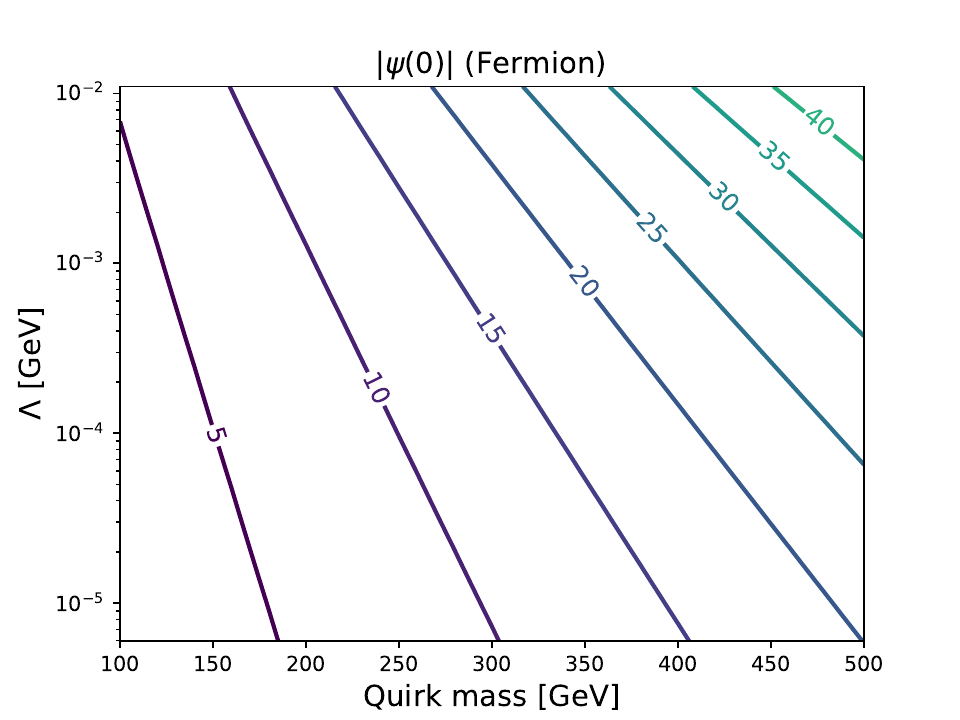}
\includegraphics[width=0.48\textwidth]{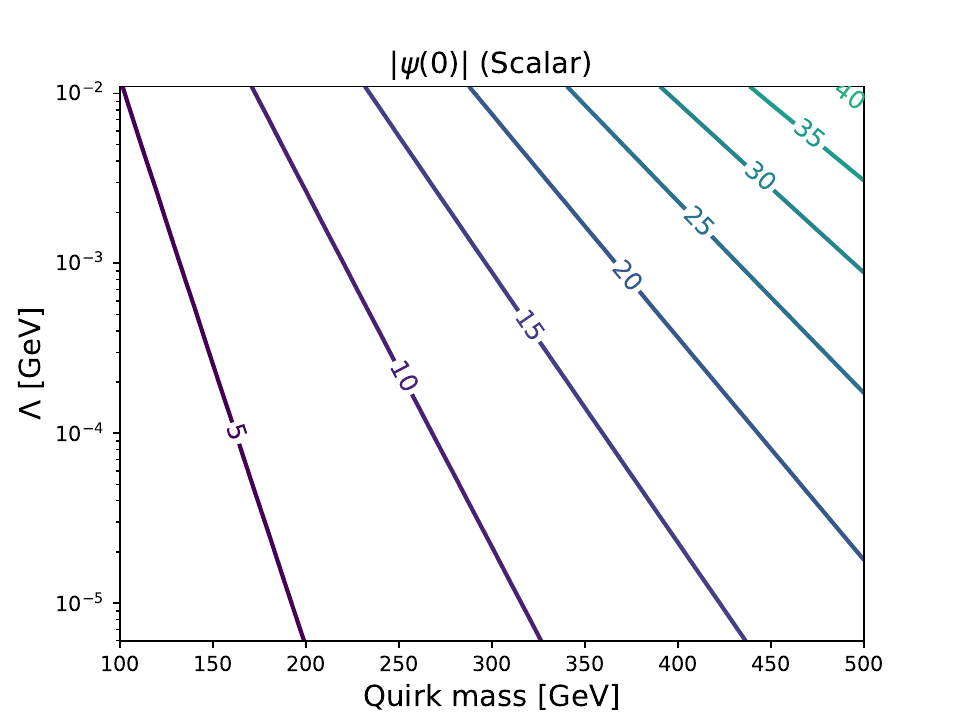}
\caption{Ground state wave function at origin.  \label{fig.wave}}
\end{figure}

Since the quirk potential depends on the quirk mass $m_{\mathcal{Q}}$ and confinement scale $\Lambda$ through the running of $\alpha_{\text{IC}}$, the absolute value of the ground state wave function at origin also depends on those values. The values of $\left|\psi(0)\right|$ on the $m_{\mathcal{Q}}$-$\Lambda$ plane are shown in Fig.~\ref{fig.wave}. 
We can observe that the $\left|\psi(0)\right|$ is increased with increasing $m_{\mathcal{Q}}$ and $\Lambda$. And the increase is more significant at larger values of $m_{\mathcal{Q}}$ and $\Lambda$. 

\begin{figure}[thbp]
\includegraphics[width=0.48\textwidth]{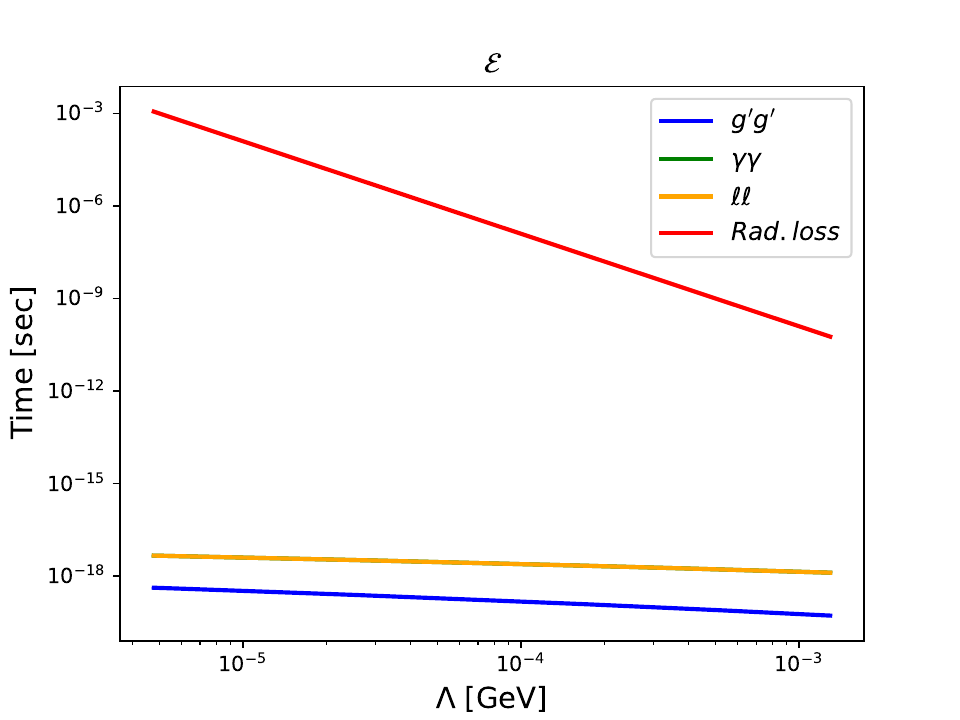}
\includegraphics[width=0.48\textwidth]{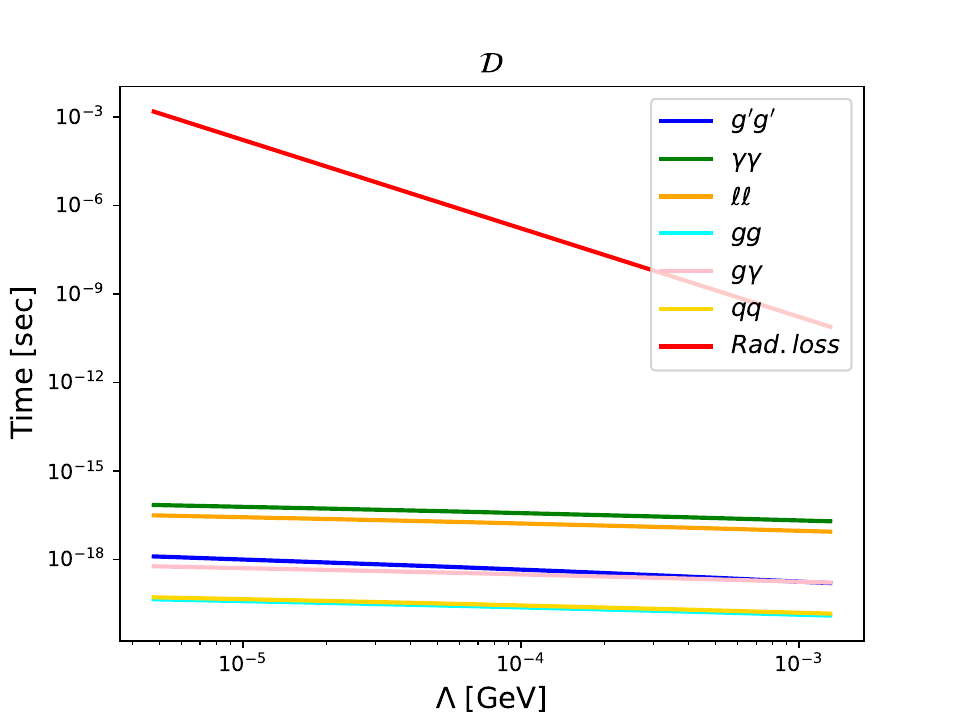}\\
\includegraphics[width=0.48\textwidth]{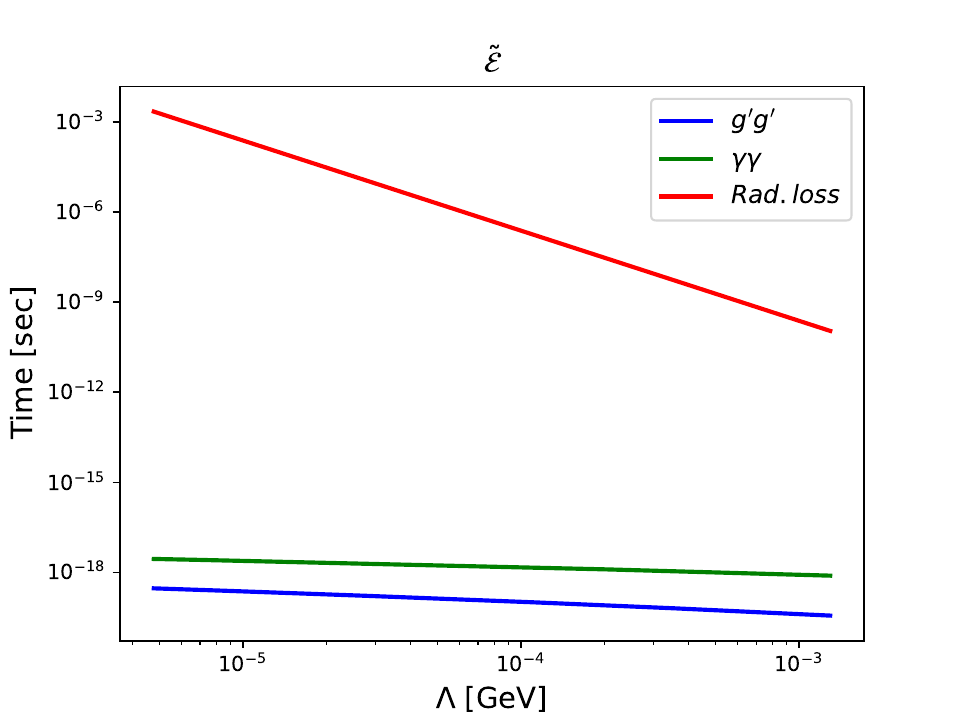}
\includegraphics[width=0.48\textwidth]{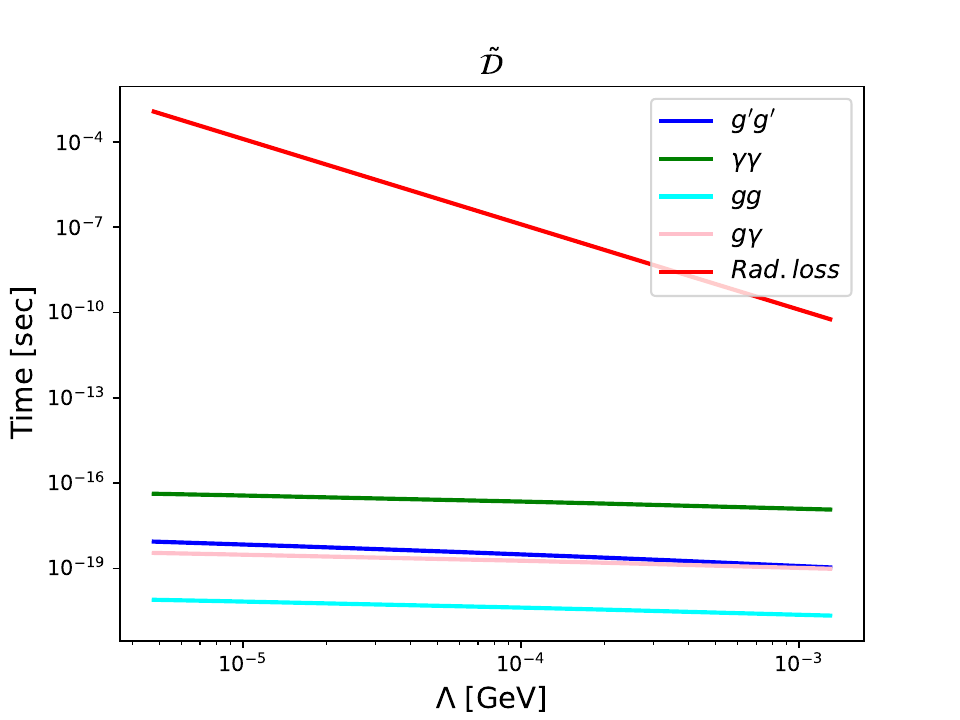}
\caption{Time scale for the energy loss through radiation and for the decay of each channel. {Relevant parameters are set as $m_{\mathcal{Q}}=100$ GeV, $\epsilon=0.1$, and $\epsilon^\prime=0$.}
 \label{fig.tquirk}}
\end{figure}

By using the quirk annihilation cross section and the wave function at origin, we can calculate the time scale ($=1/\Gamma(B \to X)$) for each decay channel. 
{The results for the $\tilde{\mathcal{D}}$, $\tilde{\mathcal{E}}$, $\mathcal{D}$, and $\mathcal{E}$ scenarios, taking into account the emission of infracolor glueballs ($\epsilon=0.1$) and disregarding QCD radiation for colored quirks ($\epsilon^\prime=0$), are presented in Fig.~\ref{fig.tquirk}.} The quirk mass $m_{\mathcal{Q}}=100$ GeV is chosen. 
The colored quirks dominantly annihilate into the gluons and/or quarks through the strong coupling, with the time scale $\sim 10^{-20}$ s. While the color-neutral quirks dominantly annihilate into the infracolor gluons through the new confining gauge interaction, with the time scale $\sim 10^{-18}$s. 
It means the quirk pair annihilates promptly once it settles into the ground state. 
The time scales for all decay channels exhibit weak dependence on the confinement scale, attributed to the wave function factor in the decay width of the bound state. 
The time scales for kinetic energy loss through radiation are also presented for comparison. 
Since the time required for energy loss depends on the initial kinetic energy which
differs event by event, the averaged value over all events in the FASER2 detector is shown for each scenario. 
Because the {$\tau^{\rm IC}_{\rm linear}$ dominate the radiation process, the time for energy loss $\propto \Lambda^{-3}$, as given in Eq.~\ref{eq:tlic}}.  
It can be seen that the time scale for energy loss is much larger than that of the annihilation for all scenarios in the parameter region of interest ($\Lambda < \mathcal{O}(1)$ MeV). In order to stimulate the signal in far detectors (with typical distance $\sim \mathcal{O}(100)$ m), the confinement scale of quirk should not be larger than $\sim$100 keV.

\subsection{A delayed annihilation scenario}

As has been discussed above, the quirk-pair bound state consisting of two quirks of the same flavor will inevitably annihilate into infracolor gluon, with typical time scale $\sim 10^{-18}$ s, leading to the prompt decay. 
In the $W_R$ scenario, there are two flavors of quirks. The annihilation of the quirk-pair bound state into infracolor gluon is forbidden due to flavor conservation. So, the lifetime of quirk-pair bound state can possibly reach $\gtrsim 10^{-6}$ s in some parameter space. 

At the LHC, a pair of $\tilde{\mathcal{V}}$-$\tilde{\mathcal{E}}$ can be produced through the s-channel $W_R$ mediation. 
In the heavy $W_R$ region, the dominant decay channel of the $\tilde{\mathcal{V}}$-$\tilde{\mathcal{E}}$ bound state is the three-body decay $\tilde{\mathcal{V}} \tilde{\mathcal{E}} \to W_R^*(\to f \bar{f}) \gamma$. The decay width of this channel is given as 
\begin{align}
\Gamma=\frac{N_f \left|\psi(0)\right|^2}{128 \pi^3 m_\mathcal{Q}^2} \int_{0}^{m_\mathcal{Q}}dE_3\int_{m_\mathcal{Q}-E_3}^{m_\mathcal{Q}}|\mathcal{M}|^2dE_1
\end{align}
with the amplitude square
\begin{align}
|\mathcal{M}|_s^2=-\frac{16 \kappa^4 g_{\text{EM}}^6 m_\mathcal{Q} N_{\text{IC}} (E_1-m_\mathcal{Q}) (E_1+E_3-m_\mathcal{Q})^2}{E_1^2 \sin^4{\theta_W} \left(4 E_1 m_\mathcal{Q}-4 m_\mathcal{Q}^2+m_{W_R}^2\right)^2}~,
\end{align} 
where $N_f=9$ is the number of the SM fermion with different flavors and colors, $\theta_W$ is the Weinberg angle, and $g_{\text{EM}}=\sqrt{4 \pi \alpha_{\text{EM}}}$. 

\begin{figure}[htbp]
\includegraphics[width=0.48\textwidth]{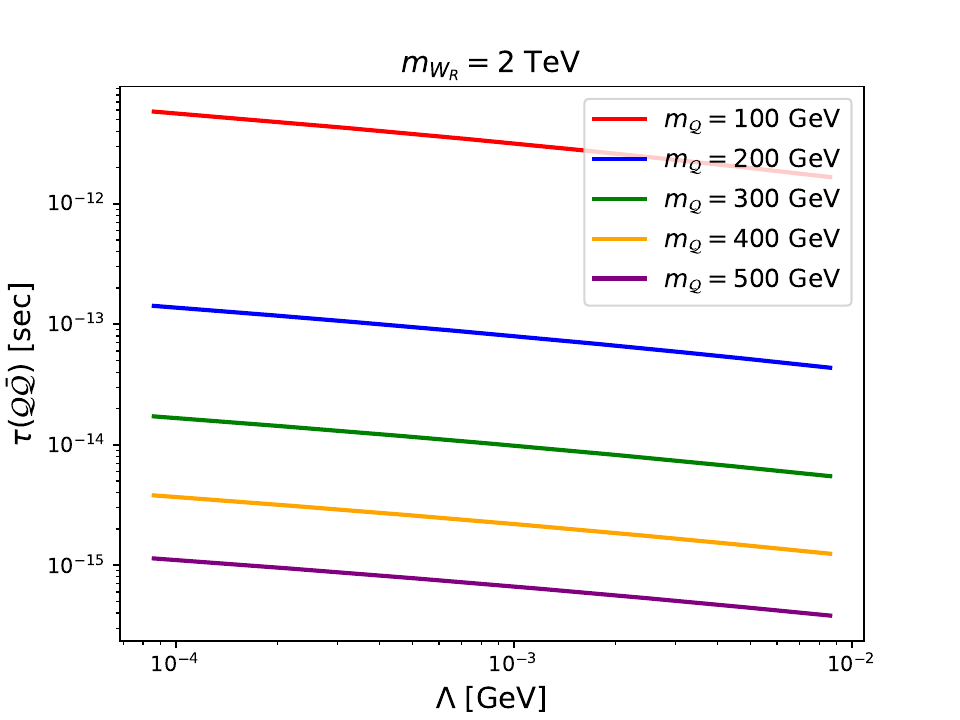}
\includegraphics[width=0.48\textwidth]{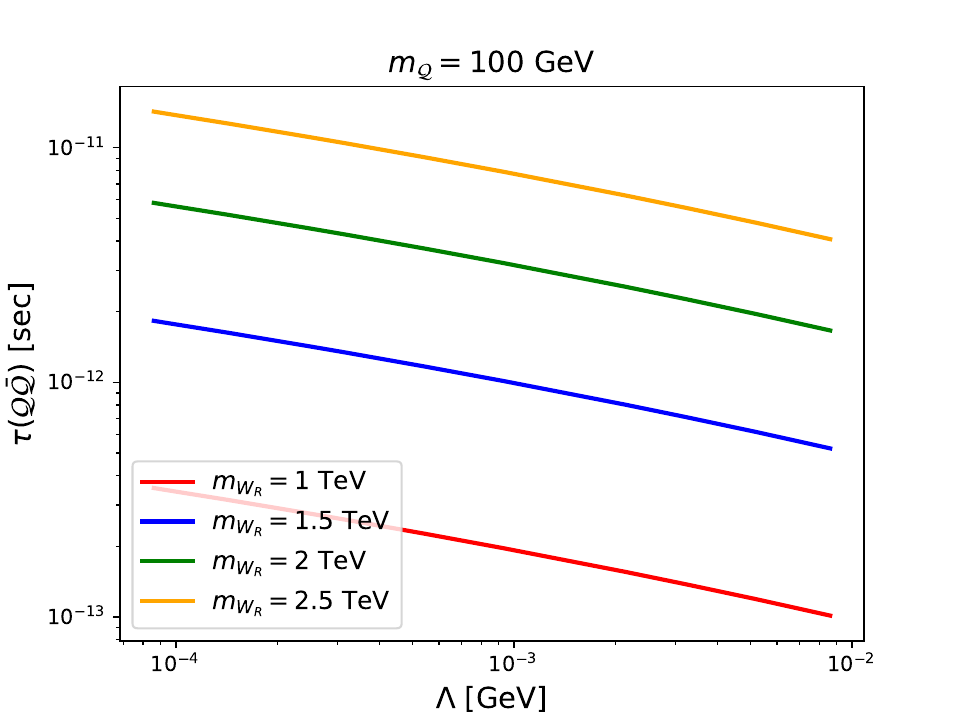}
\caption{Lifetime of the $\tilde{\mathcal{V}}$-$\tilde{\mathcal{E}}$ bound state in the $W_R$ scenario. The $\kappa=1$ has been set.  \label{fig.twr}}
\end{figure}

The Fig.~\ref{fig.twr} illustrates the lifetime dependence on the confinement scale with various quirk and the $W_R$ masses, where we have chosen $\kappa=1$. 
The dependence of the lifetime on the confinement scale originates from the wave function. 
For $\Lambda \sim 100$ keV and $m_{\tilde{\mathcal{V}}} \simeq m_{\tilde{\mathcal{E}}} \equiv m_{\mathcal{Q}}$, the lifetime of the $\tilde{\mathcal{V}}$-$\tilde{\mathcal{E}}$ bound state can be roughly estimated by 
\begin{align}
\tau = 10^{-7}~\text{s} \times (\frac{100~\text{GeV}}{m_{\mathcal{Q}}})^{5.3} \cdot (\frac{m_{W_R}}{2000~\text{GeV}})^4 \cdot (\frac{0.086}{\kappa})^4~.~ \label{eq:tani}
\end{align} 
Although smaller $\kappa$ can provide longer lived quirk bound state, the production cross section of the quirk is suppressed by a factor of $\kappa^2$. 
In order to induce promising signal at the far detectors, an appropriate parameter set needs to be chosen. 

\subsection{The time scale of $\beta$-decay}

In the $W_R$ scenario, the $\tilde{\mathcal{V}}$$\tilde{\mathcal{E}}$ bound state can undergo $\beta$ decay if there is mass difference between the two flavors of quirks. 
This decay can be approximated as the decay of the heavier quirk into the lighter one through the (off-shell) $W_R$ mediation. 
The rate of the three-body decay is
\begin{align}
\Gamma=\frac{1}{2 m_H} \frac{1}{4(2\pi)^3}\int_{E_{3}|_{\min}}^{E_{3}|_{\max}}dE_3\int_{E_{1}|_{\min}}^{E_{1}|_{\max}}|\mathcal{M}|^2dE_1
\end{align}
with the amplitude square
\begin{align}
|\mathcal{M}|_s^2=-\frac{4 N^2_{\text{IC}} g_{\text{EM}}^4 m_H^2  \left(4 E_1 E_3-2 E_1 m_H+4 E_3^2-4 E_3 m_H+m_H^2+m_L^2\right)}{ \sin^4{\theta_W} \left(-2 E_1 m_H-m_{W_R}^2+m_H^2+m_L^2\right)^2}~.~
\end{align}
The limits of integrations are given as 
\begin{align}
E_{3}|_{\min} &=0~, \\
E_{3}|_{\max} & =\frac{m_H^2-m_L^2}{2 m_H}~, \\
E_{1}|_{\min} & =\frac{\left(m_H^{2}+m_L^{2}-2 E_3 m_H \frac{\left(m_H^{2}-2 E_3 m_H-m_L^{2}\right)}{m_H^{2}-2 E_3 m_H}\right)}{2 m_H}~, \\
E_{1}|_{\max} &=\frac{m_H^2+m_L^2}{2 m_H}~.
\end{align}
In the above, $m_H$ and $m_L$ are masses of the heavy and light quirk, respectively. 
If the $\beta$ decay happens before the $\tilde{\mathcal{V}}$-$\tilde{\mathcal{E}}$ annihilation, the subsequent bound state which consists of quirk pair with the same flavor will annihilate into infracolor gluon immediately. 
As a result, the $\beta$ decay tends to reduce the signal efficiency at the far detectors. 
However, the decay width strongly depends on the mass difference between $\tilde{\mathcal{V}}$ and $\tilde{\mathcal{E}}$. 
In the $W_R$ scenario, the $\beta$ decay become irrelevant {when $\Delta m \lesssim 0.3$} GeV, which corresponds to the time scale of $\beta$ decay $\tau \gtrsim 10^{-6}$ s. 
In the following, we will simply ignore the $\beta$ decay by assuming a small mass difference.

\subsection{Evolution History}
After the quirk pair is produced, it will go through radiative energy loss and annihilation. 
The travel distance during the radiation can be simply calculated as $d_{\rm rad}= \beta \times t^{\rm lab}_{\rm rad}$ with $t^{\rm lab}_{\rm rad}$ given in Eq.~(\ref{tlabrad}).
The decay length of the quirk pair in the laboratory frame $d_{\rm anni}= \beta \gamma c\tau_{\mathcal{Q} \bar{\mathcal{Q}}}$ where $\tau_{\mathcal{Q} \bar{\mathcal{Q}}} = 1/ \Gamma^{\rm tot}_{\mathcal{Q} \bar{\mathcal{Q}}}$ is the mean proper decay (annihilation) time, $\beta = p/E$ is velocity, and $\gamma = E/m$ is the boost factor. 
Combining those stages, we can calculate the total signal-event rates at the detectors as
\begin{align}
N=\mathcal{L} \times  \int_{\Delta \Omega} d\Omega_{\mathcal{Q} \bar{\mathcal{Q}}} \int_{p_{\mathcal{Q} \bar{\mathcal{Q}}}} dp_{\mathcal{Q} \bar{\mathcal{Q}}} \cdot \frac{d \sigma_{pp\to \mathcal{Q} \bar{\mathcal{Q}}} }{dp_{\mathcal{Q} \bar{\mathcal{Q}}} d\Omega_{\mathcal{Q} \bar{\mathcal{Q}}}}\mathcal{P}(p_{\mathcal{Q} \bar{\mathcal{Q}}})~, \label{eq:nquirkdet}
\end{align}
where the $\mathcal{L}$ is the integrated luminosity, $\sigma_{pp\to \mathcal{Q} \bar{\mathcal{Q}}}$ is the quirk production cross section, and $\Delta \Omega$ is the geometric cross-section of the active detector volume. 
The $\mathcal{P}(p_{\mathcal{Q} \bar{\mathcal{Q}}})$ describe the probability of the quirk pair (with total momentum $p_{\mathcal{Q} \bar{\mathcal{Q}}}$) annihilating inside the detector. 
We consider two different kinds of quirk signals. The first one is the charged quirk pair traveling through the detector and leaving the charged track that can be distinguished from the backgrounds. The $\mathcal{P} (p_{\mathcal{Q} \bar{\mathcal{Q}}})$ for this case is given by 
\begin{align}
\mathcal{P}_{\mathcal{Q} \bar{\mathcal{Q}}}=\begin{cases}
		1 & \text{ if } d_{\rm rad}> L_{\rm in} \\ 
		\int^{\infty}_{L_{\rm in}-d_{\rm rad}}\frac{1}{d_{\rm anni}}e^{-\frac{L}{d_{\rm anni}}}dL & \text{ if } d_{\rm rad} \le L_{\rm in}  
	\end{cases}~.~ \label{eq:prob1}
\end{align}
In the second case, we require the quirk pair to annihilate inside the detector. This not only enables the reconstruction of the secondary vertex but also helps in the identification of annihilation final states as well as in the measurement of the quirk pair momentum. However, this restriction will limit the number of signal events, especially in the case of long-lived quirk bound state when $\Lambda$ is relatively small. The $\mathcal{P} (p_{\mathcal{Q} \bar{\mathcal{Q}}})$ is given by 
\begin{align}
\mathcal{P}_{\mathcal{Q} \bar{\mathcal{Q}}}=\begin{cases}
		0 & \text{ if } d_{\rm rad}> L_{\rm out} \\ 
		\int^{L_{\rm out}-d_{\rm rad}}_{\max \{ 0,L_{\rm in}-d_{\rm rad} \}}\frac{1}{d_{\rm anni}}e^{-\frac{L}{d_{\rm anni}}}dL & \text{ if } d_{\rm rad} \le L_{\rm out}  
	\end{cases}~.~  \label{eq:prob2}
\end{align}
In the Eqs.~(\ref{eq:prob1}), (\ref{eq:prob2}) above, the $L_{\rm in}$ and $L_{\rm out}$ denote the travel distances for quirk to reach and leave the detector volume, respectively. 
We note that the $L_{\rm in}$ and $L_{\rm out}$ depend on the direction of the quirk pair momentum. In practice, the Eq.~(\ref{eq:nquirkdet}) is calculated numerically by simulating the quirk production events with \texttt{MG5\_aMC@NLO}. So, the $L_{\rm in}$, $L_{\rm out}$ and $\mathcal{P}_{\mathcal{Q} \bar{\mathcal{Q}}}$ are calculated on each event.

\section{Long-lived particle detectors at the LHC} \label{sec:det}

There have been a number of proposed and installed far detectors in the vicinity of IP of the LHC. In this work, we perform the numerical analysis taking into account the FASER2, SND@LHC, ANUBIS, FACET, and MATHUSLA detectors. They are sensitive to the signature of long-lived quirk pair being either traveling through the detectors or decaying inside their fiducial volumes. 

FASER (ForwArd Search ExpeRiment) is located 480 m downstream from the ATLAS IP. It is a cylindrical detector with a central axis located on the $z$-axis. Its decay volume has a diameter of 0.2 m and a length of 1.5 m. It has reported the first observation of collider neutrino events~\cite{FASER:2023zcr} and the results of the dark photon search~\cite{FASER:2023tle}. The detector will be upgraded to FASER2 with a larger decay volume (2 m diameter and 5 m length), providing better acceptance to LLPs. 
The SND@LHC (The Scattering and Neutrino Detector at the LHC) detector is located 480 m downstream from the ATLAS IP, has a length of 2.47 m along the $z$-axis, and is slightly off-axis. It consists of a vertex detector and electromagnetic calorimeter followed by a hadronic calorimeter and a muon identification system. The detector is capable of identifying all three neutrino flavors with high efficiency. The direct observation of muon neutrino interactions with the detector was reported in Ref.~\cite{SNDLHC:2023pun}.
ANUBIS (AN Underground Belayed In-Shaft search experiment) is an off-axis cylindrical detector designed for searching neutral LLPs with travel distances larger than 5 m. It is proposed to occupy the PX14 installation shaft of the ATLAS experiment, with four tracking stations. The central axis of the cylinder is 14 m from the IP. The detector may be divided into three segments with the length of each segment roughly equal to 18.7 m.
FACET (Forward-Aperture CMS ExTension) is located 101 m downstream from the CMS IP. It has a decay volume of length of 18 m and a radius of 0.5 m, followed by an 8-meter-long region instrumented with various particle detectors. A unique feature among the LHC experiments is that the decay volume is at a high vacuum, eliminating most background from particle interactions inside a 14 $\text{m}^3$ fiducial region. Although it has a relatively small radius, its angular coverage is larger than other detectors due to its closeness to the IP. 
MATHUSLA (MAssive Timing Hodoscope for Ultra-Stable neutraL pArticles) is located 70 m downstream from the CMS IP, slightly off-axis and partly above ground. It has a decay volume of size $100~\text{m} \times 100 ~\text{m} \times 25~\text{m}$, a floor detector as a tracking veto to reject back-scattered charged particles and cosmic muon, and a double-layer tracker at ground level to enhance the precision of track reconstruction. 

\begin{figure}[htbp]
\includegraphics[width=0.85\textwidth]{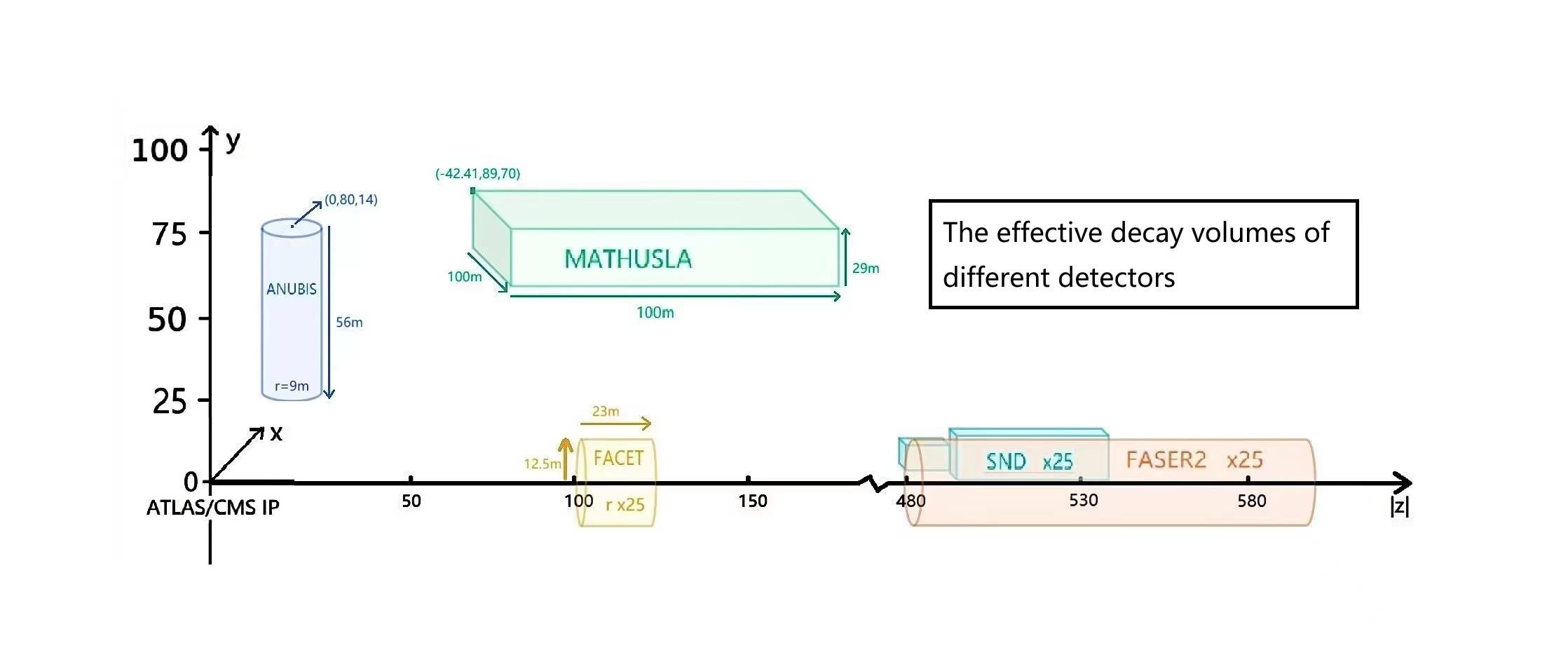}
\caption{The schematic diagram illustrates the locations and effective volumes of the detectors. The SND@LHC and FASER2 detectors are enlarged by a factor of 25 for better visibility. For the FACET detector, only the radius is scaled up by the same factor.\label{fig.detpos}}
\end{figure}

\begin{table}[tbp]
	\centering 
	\begin{tabular}{|c|c|c|} \hline
	Detector  & Effective volume \\  \hline
FACET  & $z \in [101,124]~\text{m}, \quad r = \sqrt{x^2 + y^2} \in [0,0.5]~\text{m}$ \\ \hline
ANUBIS &  $y \in [24,80]~\text{m}, \quad r=\sqrt{x^2 + (z-14)^2} \in [0,9]~\text{m}$ \\ \hline
FASER2  & $z \in [480,485]~\text{m}, \quad r=\sqrt{x^2 + y^2} \in [0,1]~\text{m}$ \\ \hline
\multirow{2}*{SND@LHC}  & section 1: $x \in [-0.47, -0.08]~\text{m}, \quad y \in [0.155, 0.545]~\text{m}, \quad z \in [-480, -480.56]~\text{m}$ \\
              & section 2: $x \in [-0.737, 0.03]~\text{m}, \quad y \in [0.064, 0.674]~\text{m}, \quad z \in [-480.56, -482.47]~\text{m}$ \\ \hline
MATHUSLA  &  $x \in [-42.41, 57.59]~\text{m}, \quad y \in [60, 89]~\text{m}, \quad z \in [70, 170]~\text{m}$ \\ \hline
      \end{tabular}
      \caption{The effective volume for each detector. \label{tab:dets}}
 \end{table}

To obtain an intuitive overview of the positions and sizes of all the detectors mentioned above, we present a schematic diagram in Fig.~\ref{fig.detpos}. For better illustration, we have magnified some dimensions of the detectors. We also provide the specific parameters for the effective detector volumes used in our simulation in Tab.~\ref{tab:dets}.

\section{Experimental sensitivities to quirk} \label{sec:exc}

This work explores the parameter space with confinement scale $\Lambda \gtrsim  3$ keV and quirk mass $m_{\mathcal{Q}}$ ranging from $100~\text{GeV}$ to $2000~\text{GeV}$. In this case, the quirk pair oscillates with a typical amplitude of $L \lesssim 1$ mm, which is much smaller than the detector sizes. Therefore, the quirk-pair system can be treated as a single LLP. To produce signals at far detectors, the quirk-pair system must have a momentum that points to the effective detector volumes. Moreover, the quirk pair must have a long enough lifetime to either travel through the detector or decay inside it.

\begin{figure}[htbp]
\includegraphics[width=0.24\textwidth]{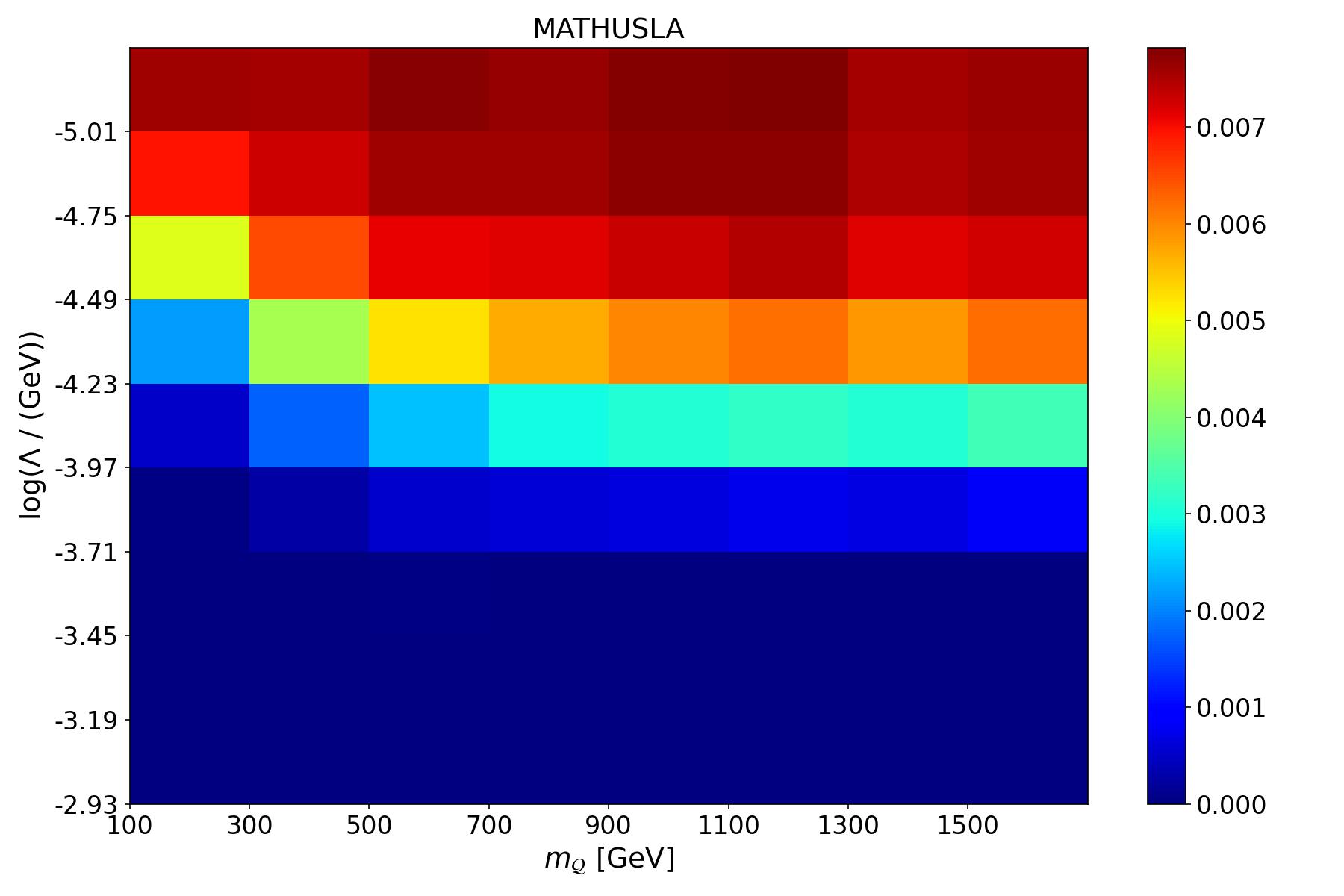}
\includegraphics[width=0.24\textwidth]{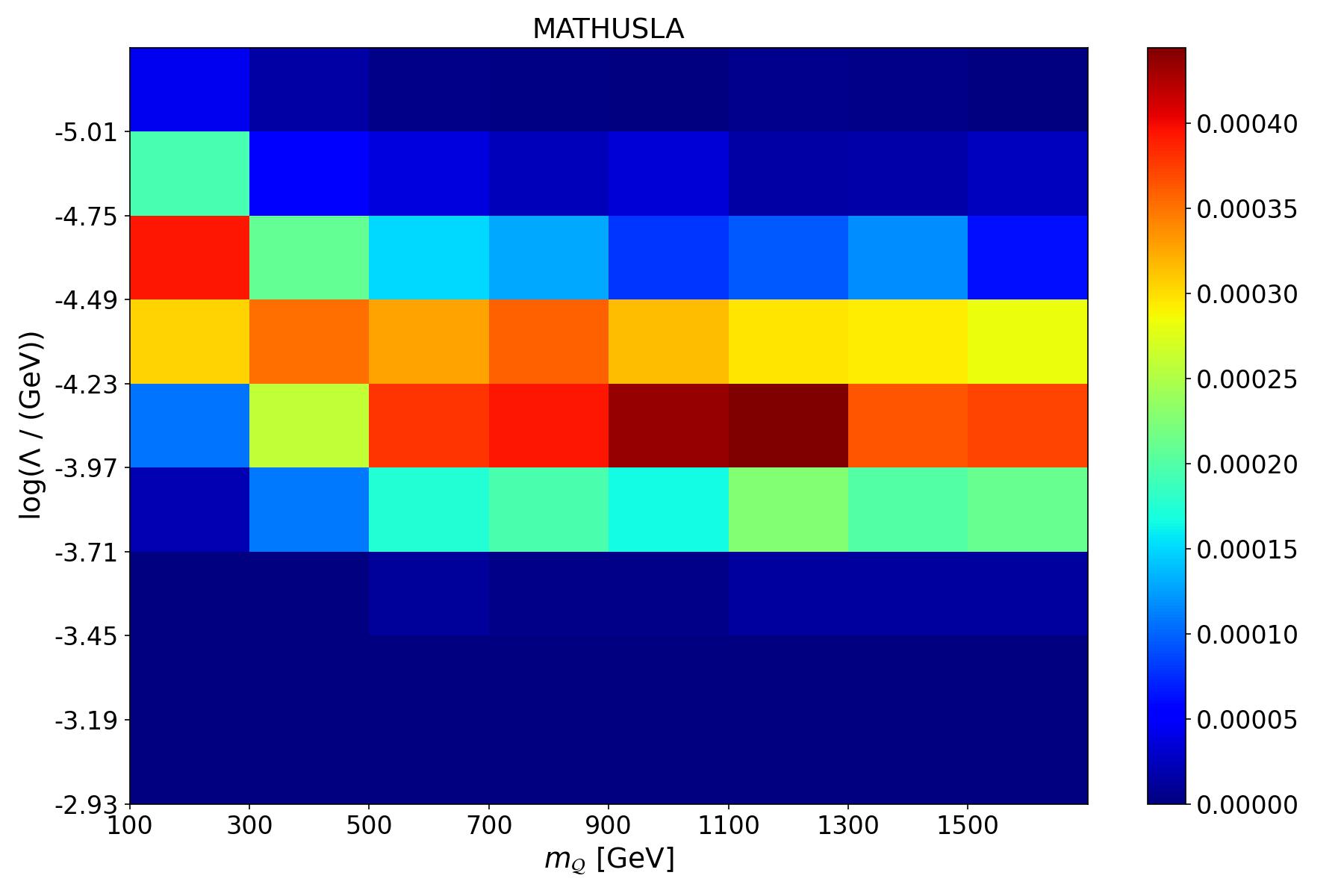}
\includegraphics[width=0.24\textwidth]{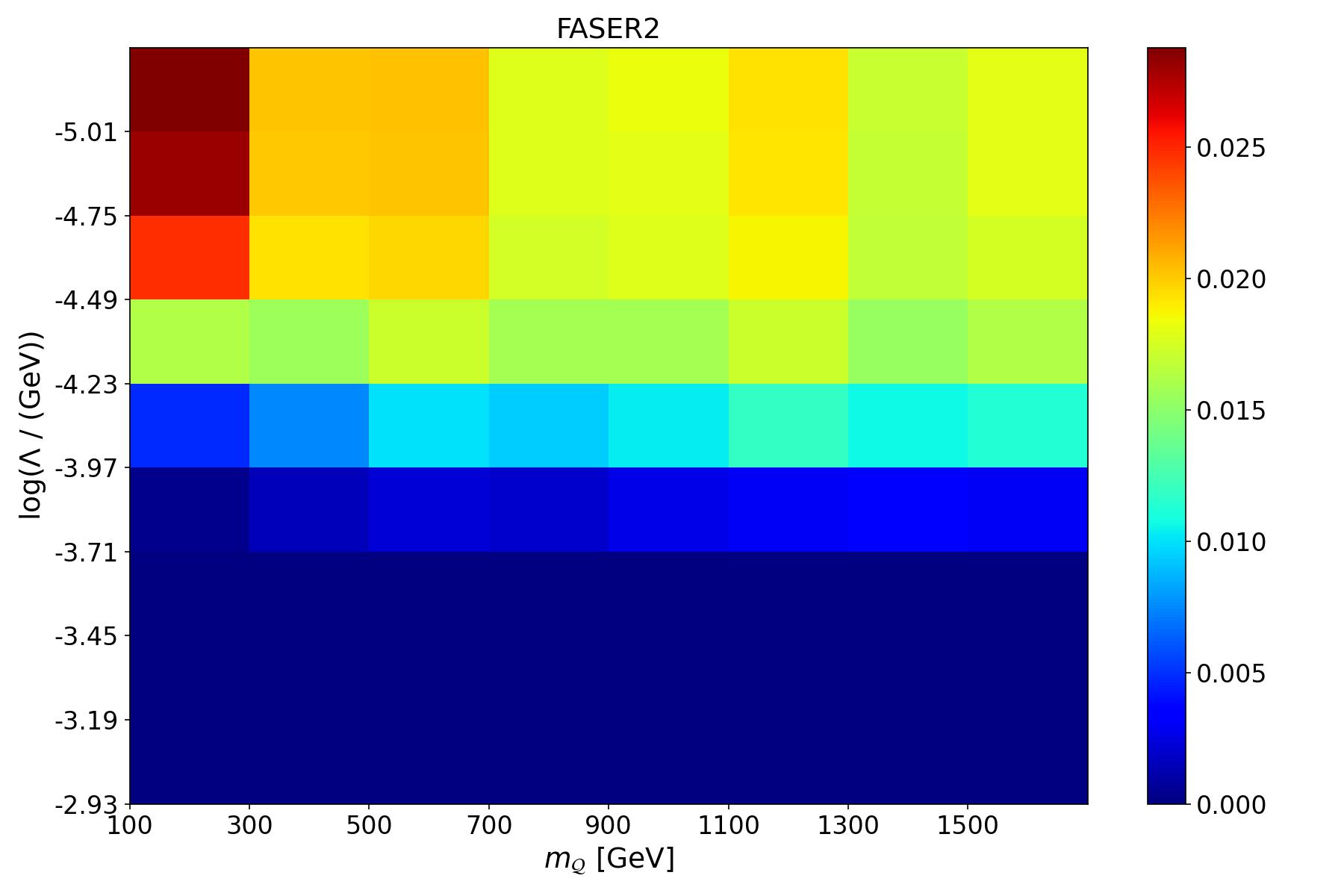}
\includegraphics[width=0.24\textwidth]{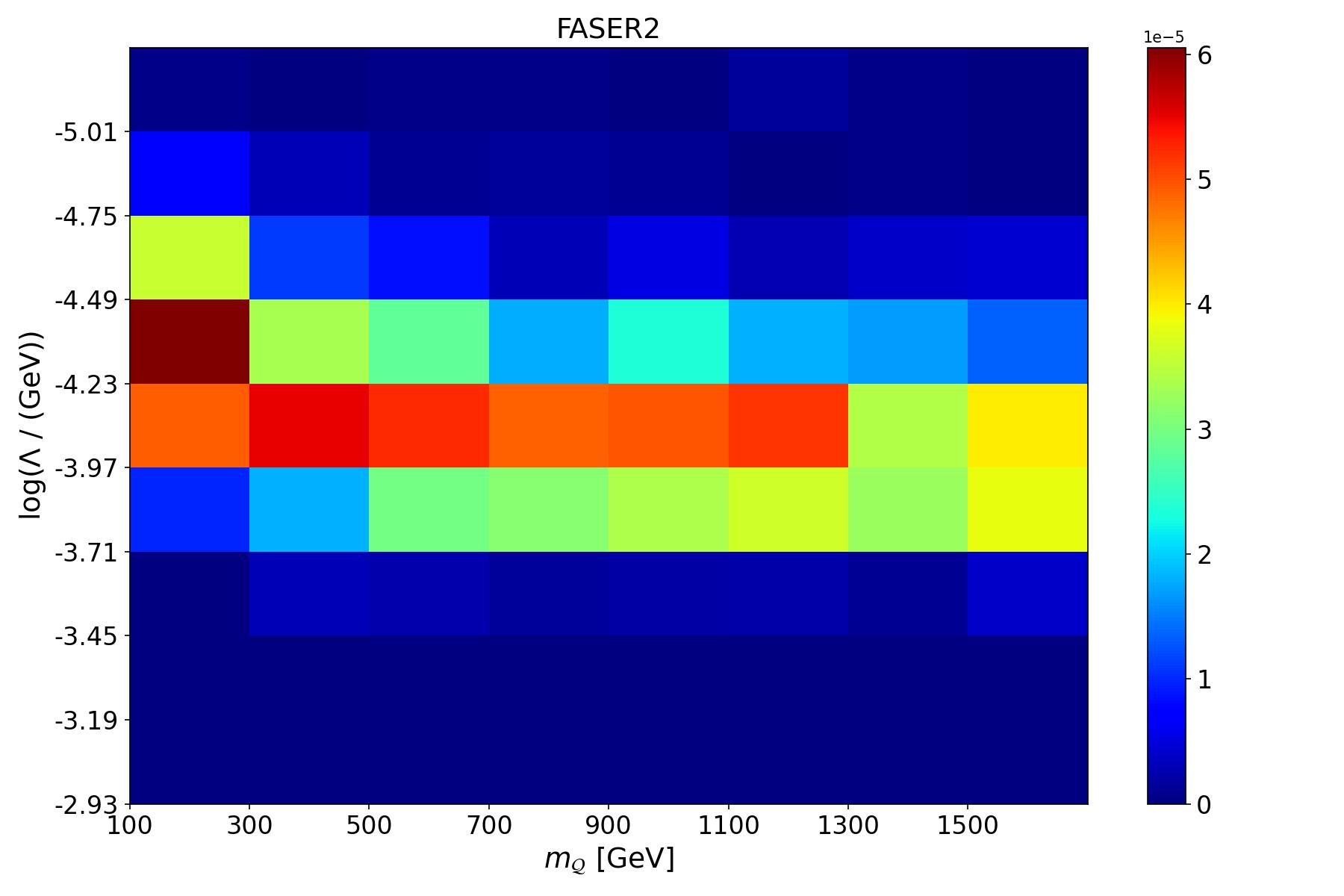} \\
\includegraphics[width=0.24\textwidth]{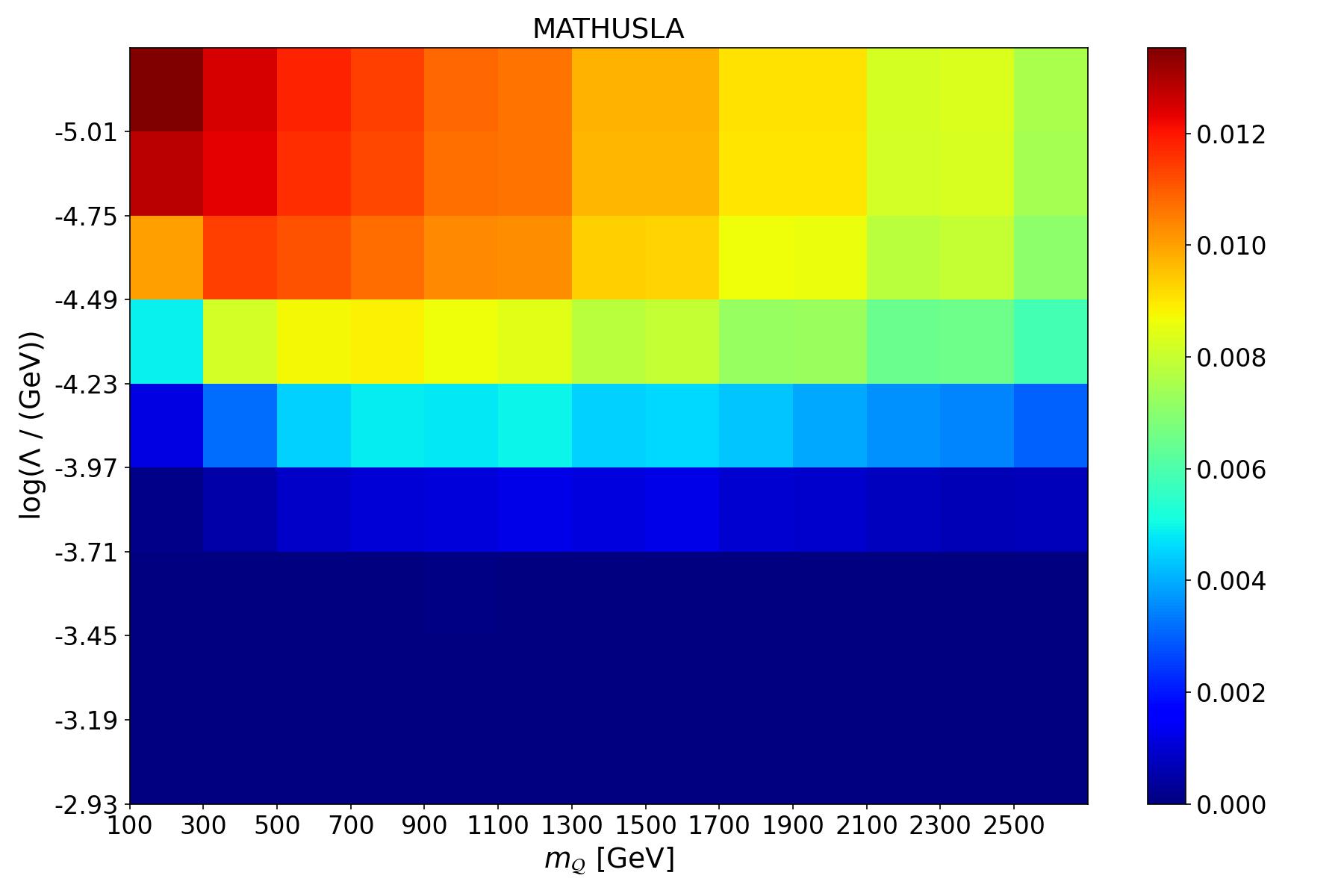}
\includegraphics[width=0.24\textwidth]{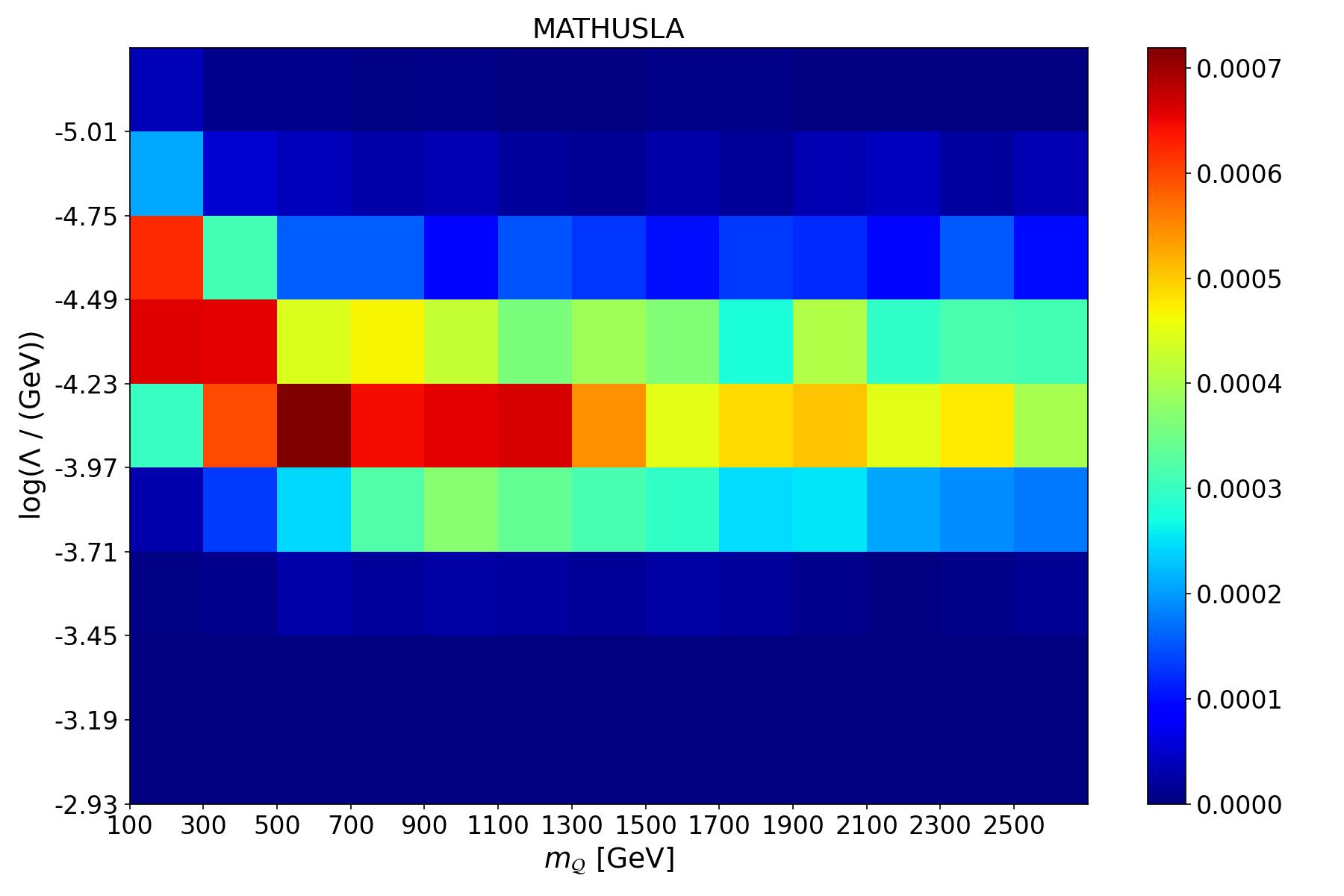} 
\includegraphics[width=0.24\textwidth]{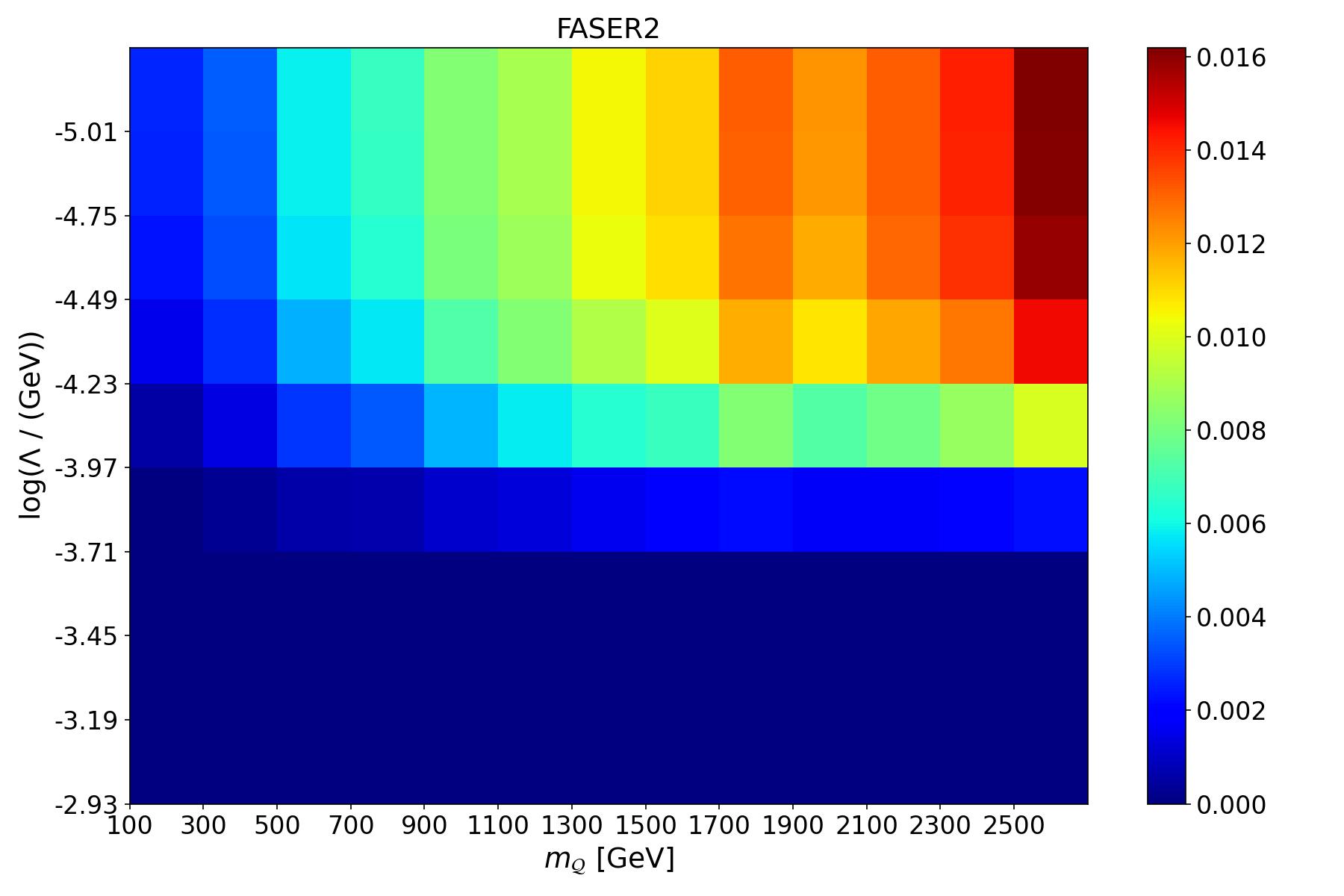}
\includegraphics[width=0.24\textwidth]{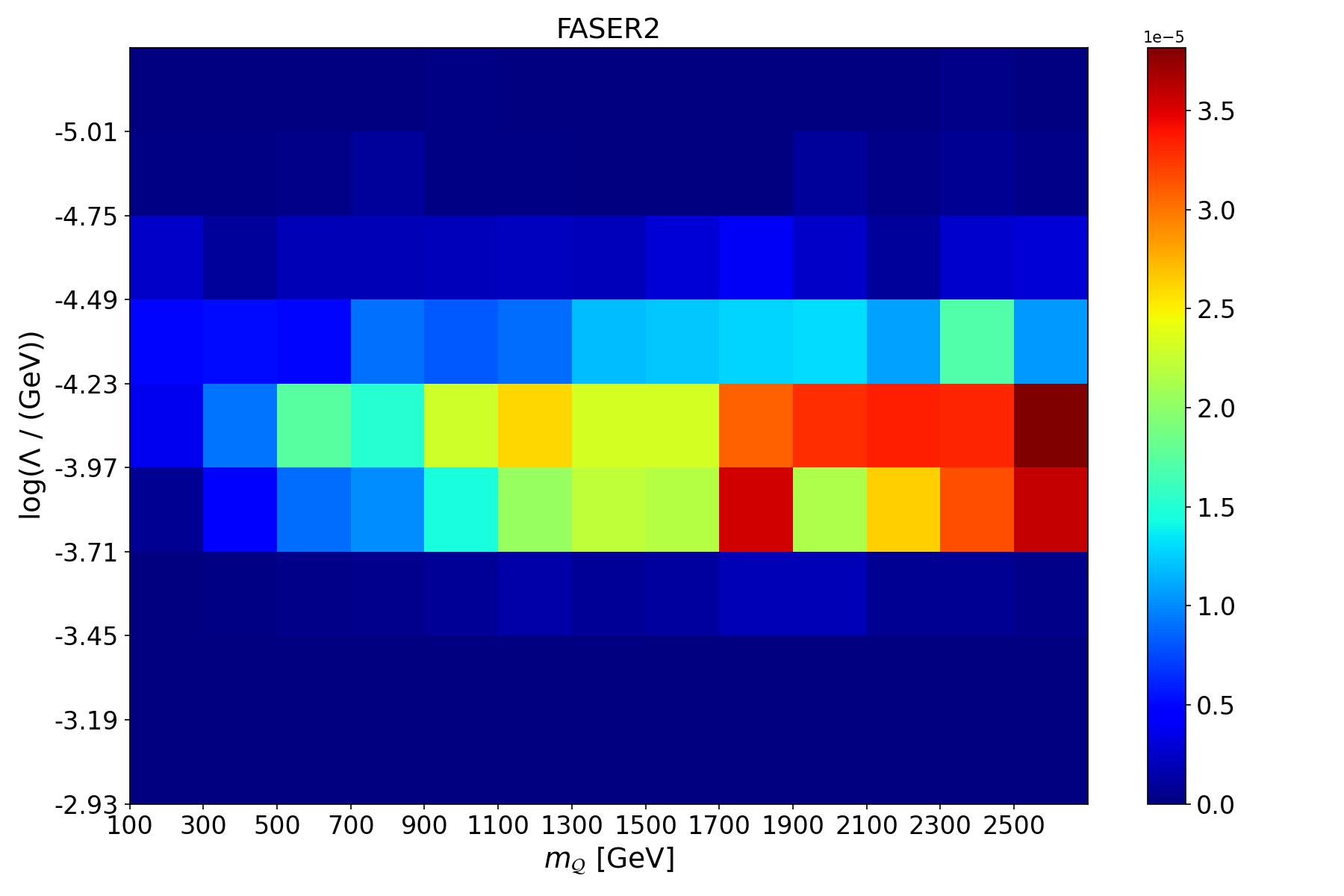} 
\caption{The panels show the fractions of quirk events, as a function of quirk mass and confinement scale, that have different signatures at MATHUSLA and FASER2. The upper panels show the results for $\mathcal{E}$, while the lower panels show the results for $\mathcal{D}$. From left to right: the fraction of events that travel through the MATHUSLA, decay inside MATHUSLA, travel through the FASER2, and decay inside FASER2. {Relevant parameters are set as $\epsilon=0.1$ and $\epsilon^\prime=0$.}
 \label{fig.effs}}
\end{figure}

{Fig.~\ref{fig.effs} presents the fractions of quirk events, categorized by quirk mass and confinement scale, demonstrating whether quirk pairs either pass through or decay within the MATHUSLA and FASER2 detectors. This analysis incorporates the emission of infracolor glueballs, with $\epsilon$ set at 0.1, and excludes QCD radiation from colored quirks by setting $\epsilon^\prime$ to 0.}
For the single quirk scenarios, the lifetime of the quirk pair is mainly determined by the period of radiative energy loss. Since both the MATHUSLA and FASER2 detectors are placed at $\sim 200-500$ m from the IP, the confinement scale should not be larger than $\sim 0.3$ MeV. This leads to the cut-off for the signal efficiency in all plots. 
The quirks in $\mathcal{E}$ and $\mathcal{D}$ scenarios have different kinematic properties as shown in Fig.~\ref{fig.xsec}. 
There are higher fractions of $\mathcal{E}$ events in the forward direction for lighter quirk. So, for lighter $\mathcal{E}$ the FASER2 (detecting forward propagating events) efficiency is higher while MATHUSLA (detecting transverse propagating events) efficiency is lower. The features become opposite for the $\mathcal{D}$ quirk. 
As for the efficiency change along the vertical direction, we can find that the traveling through signal efficiency is increased with decreasing confinement scale, {\it i.e.} longer lifetime. On the other hand, the efficiency of decay inside the detector is maximal only for a certain value of $\Lambda$, which varies in different quirk scenarios and at different detectors.

\begin{figure}[htbp]
\includegraphics[width=0.24\textwidth]{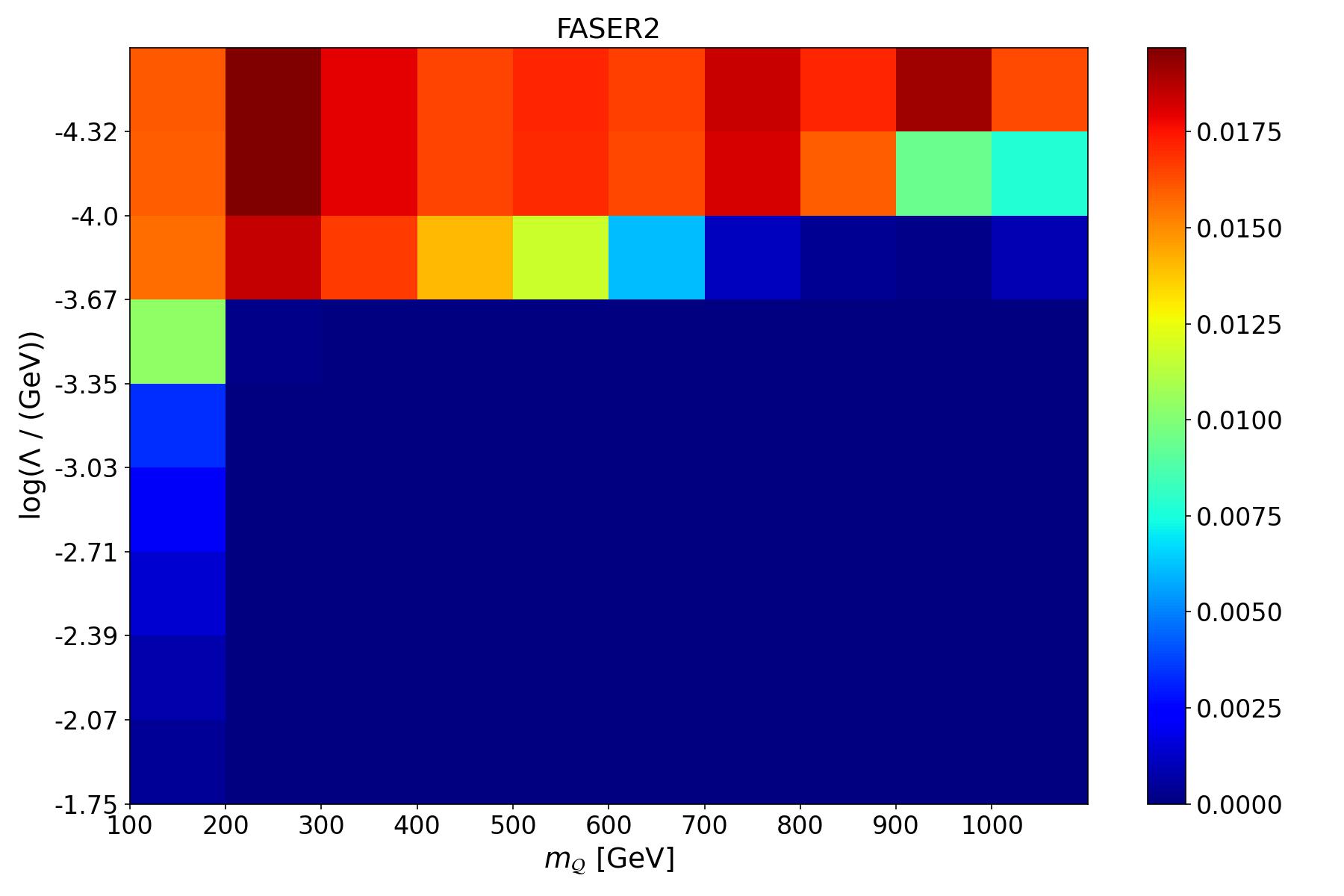}
\includegraphics[width=0.24\textwidth]{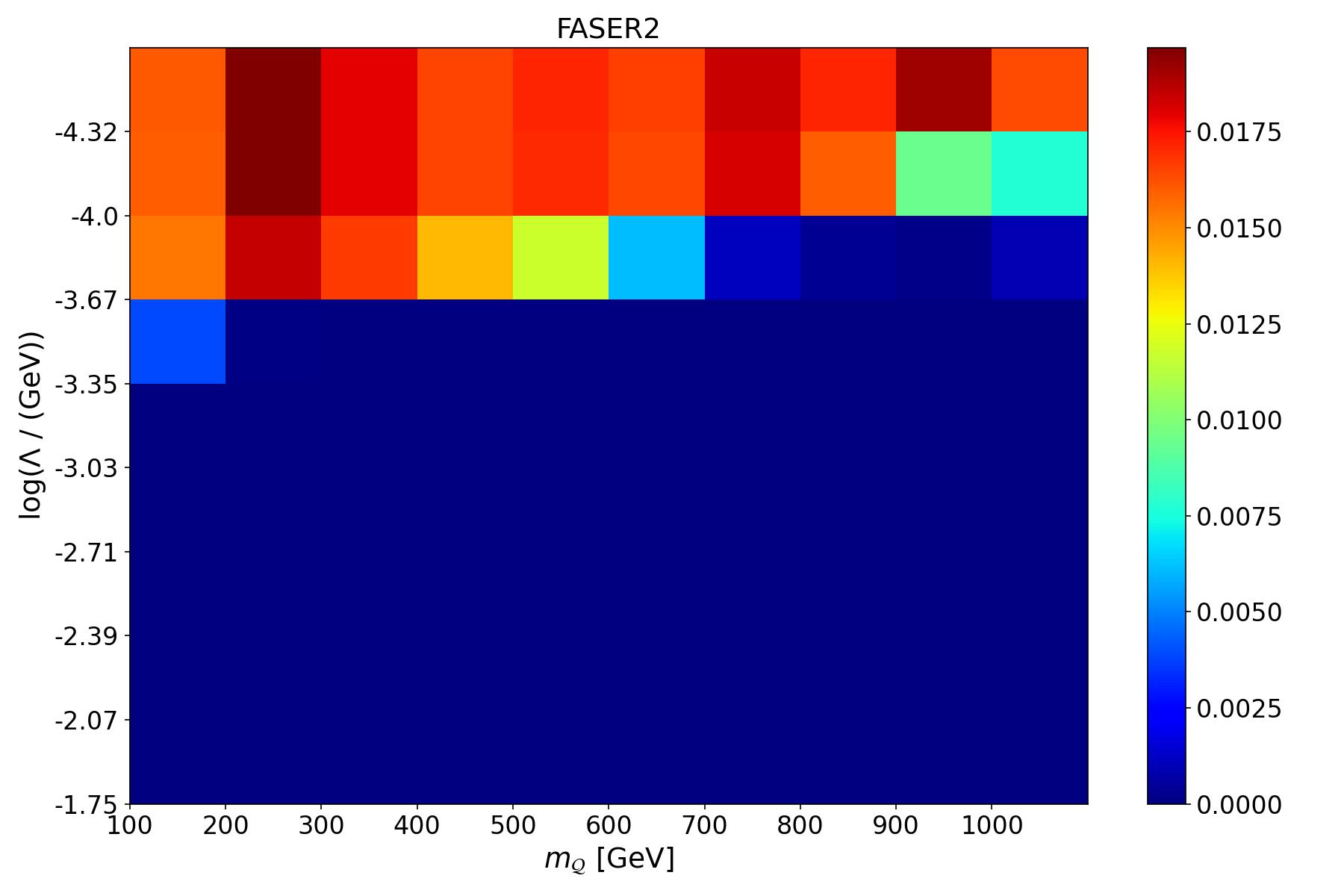}
\includegraphics[width=0.24\textwidth]{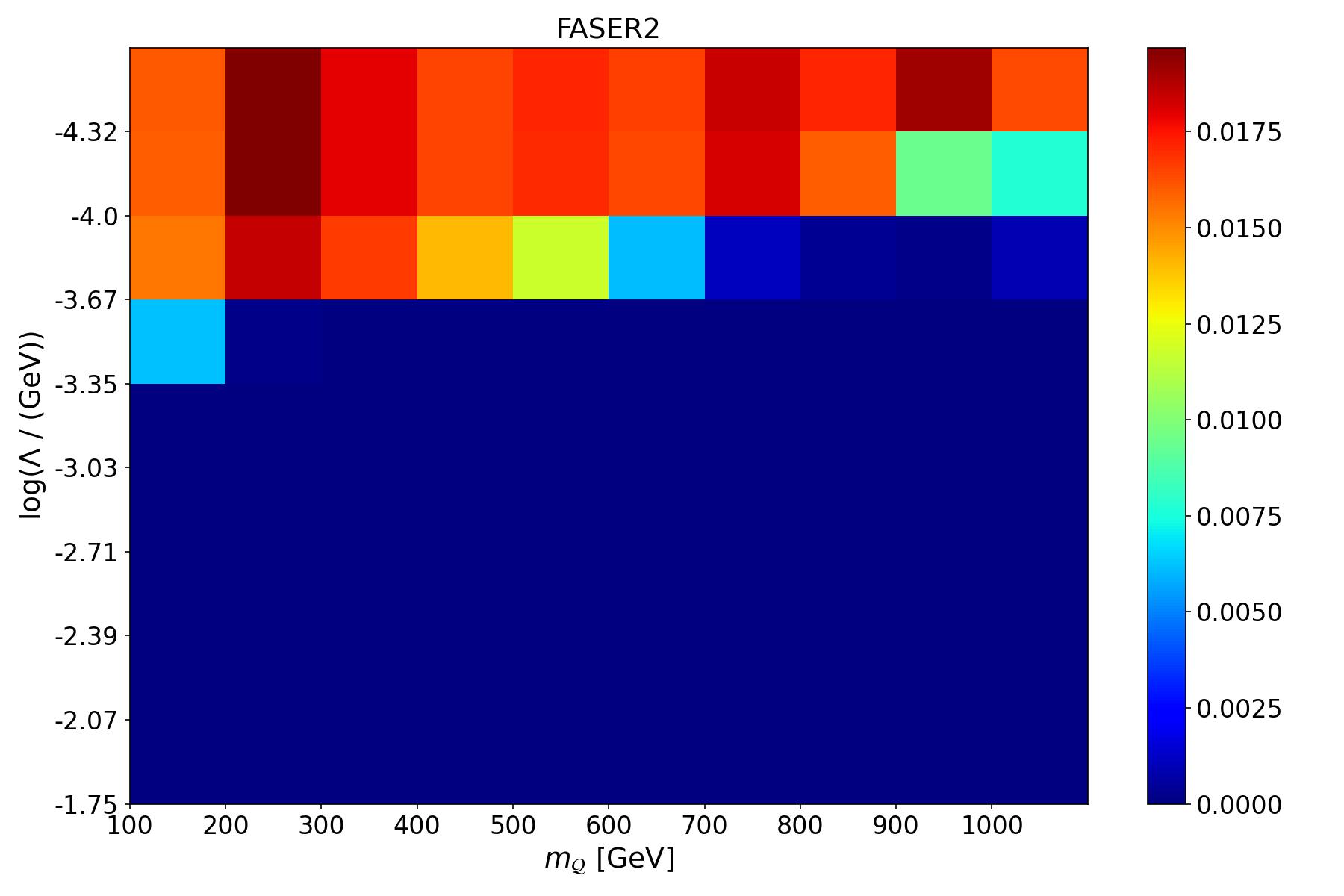}
\includegraphics[width=0.24\textwidth]{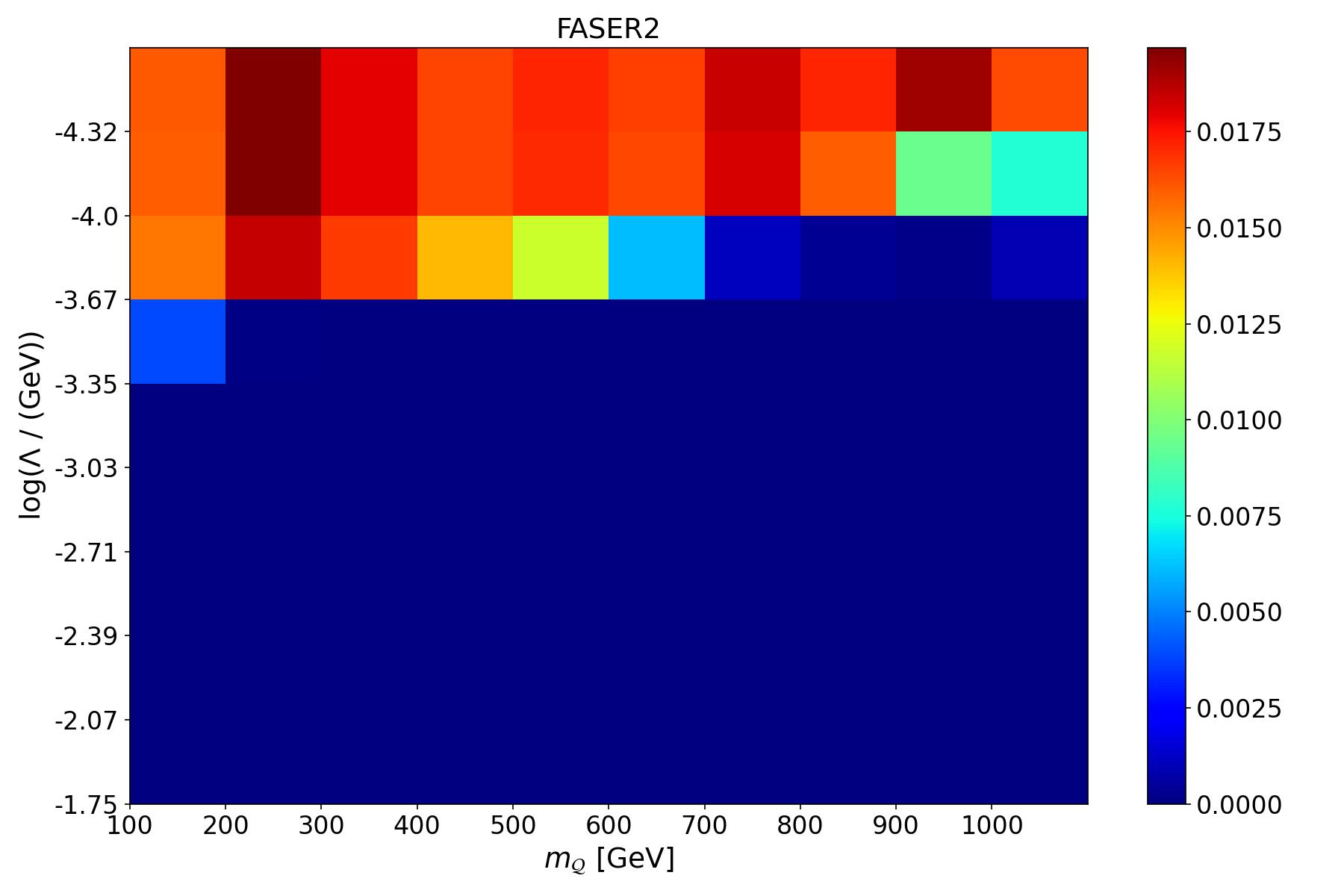} 
\caption{The panels show the fraction of quirk events in the $W_R$ scenario, as a function of quirk mass and confinement scale, that can travel through the FASER2. The $W_R$ mass has been chosen as 2 TeV. From left to right: $\kappa=0.08$ and considering time for annihilation, $\kappa=0.08$ and assuming annihilation is prompt, $\kappa=0.12$ and considering time for annihilation, $\kappa=0.12$ and assuming annihilation is prompt, respectively.
\label{fig.effswr}}
\end{figure}

Fig.~\ref{fig.effswr} shows the fraction of quirk events in the $W_R$ scenario that can travel through the FASER2 detector. 
In order to demonstrate the effects of non-negligible quirk annihilation time in this scenario, the results for the case where the quirk pair annihilation is assumed to be prompt are also presented. Comparing the first panel with the second one, we can find that the finite quirk annihilation time can help to increase the signal efficiency in the large $\Lambda$ region, where the energy loss time is short $c\tau < 100$ m. We should note that the annihilation time is dramatically reduced when the quirk mass is increasing, as given in Eq.~(\ref{eq:tani}). So, the efficiency cannot be regained for heavier quirk $m_{\mathcal{Q}} \gtrsim 200$ GeV. 
For confinement scale smaller than $\sim 10^{-3.67}(\sim 2 \times 10^{-4})$ GeV, the period of energy loss is always dominating, and the efficiency is not relevant to the annihilation. 
Each quirk is produced with total energy around half of $W_R$ mass, a heavier quirk corresponds to lower kinetic energy, thus with a shorter period of energy loss. 
As a result, the signal efficiency is reduced with increasing quirk mass for $\Lambda \sim 10^{-3.67}$ GeV. 
In the third and fourth panels, the corresponding results for larger $\kappa=0.12$ are shown. 
The quirk pair annihilation width is proportional to the $\kappa^4$. So, the annihilation time is reduced by a factor of $(0.12/0.08)^4 \sim 5$. Compared with the first and second panels, the change of efficiency is significant only in the region with $\Lambda > 10^{-3.35}(\sim  4.5 \times 10^{-4})$ GeV and $m_{\mathcal{Q}} \lesssim 200$ GeV. 

The dominant sources of background at those far detectors are the radiative processes associated with muons coming from the LHC IPs. 
For example, at the LHC Run-3, the expected flux of muon with energy larger than 100 GeV is around $0.2~\text{cm}^2~\text{s}^{-1}$ in the FASER2 detector. 
In neutral LLP searches, such a muonic background can be easily suppressed by the veto scintillator system, which is placed at the front of the detectors. 
On the other hand, the quirk carries an electric charge and enters from the front of detectors, the charge particle veto can no longer be applied. 
The ``decay inside" quirk signature can have several energetic and charged particles in the final state, with the decay vertex inside the effective detector volume. This feature is unique for new particles with TeV scale mass at those far detectors and can be easily identified. 
The ``go through" quirk signature is featured by a straight track with high ionization energy loss.
First, the quirk oscillation amplitude is unresolvable by the track system when $\Lambda \gtrsim 5$ keV. The quirk-pair system is electric neutral so that its trajectory is unbended by the Lorentz force. This gives a straight track which can be traced back to the IP. 
Second, the quirk is much heavier and moves slower than background muons. The ionization energy loss of quirk pair in the silicon strip is much larger than that of muon. 
Third, some of those far detectors are capable of providing timing information when the quirk hits the scintillators. The delayed arrival time at those scintillators as well as the time difference for quirk reach different scintillators can help to distinguish the signal from the muon background. 
Designing specific cuts to separate quirk tracks from muon tracks requires a detailed simulation of the detector environment and configuration, which is beyond the scope of the current work. 
In the following, we present the exclusion bounds at 95\% confidence level (CL) with 3 signal events, assuming zero background. Such a strategy has been widely adopted in the literature~\cite{FASER:2023tle,Beltran:2023ksw}.

\subsection{The exclusion bounds}

\begin{figure}[thbp]
\includegraphics[width=0.48\textwidth]{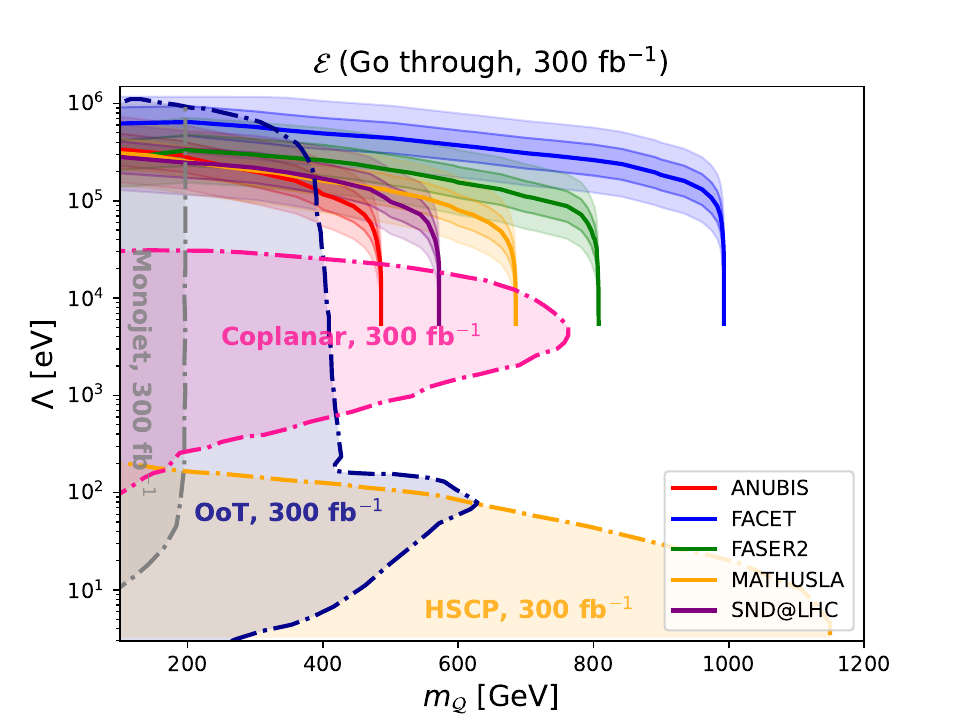}
\includegraphics[width=0.48\textwidth]{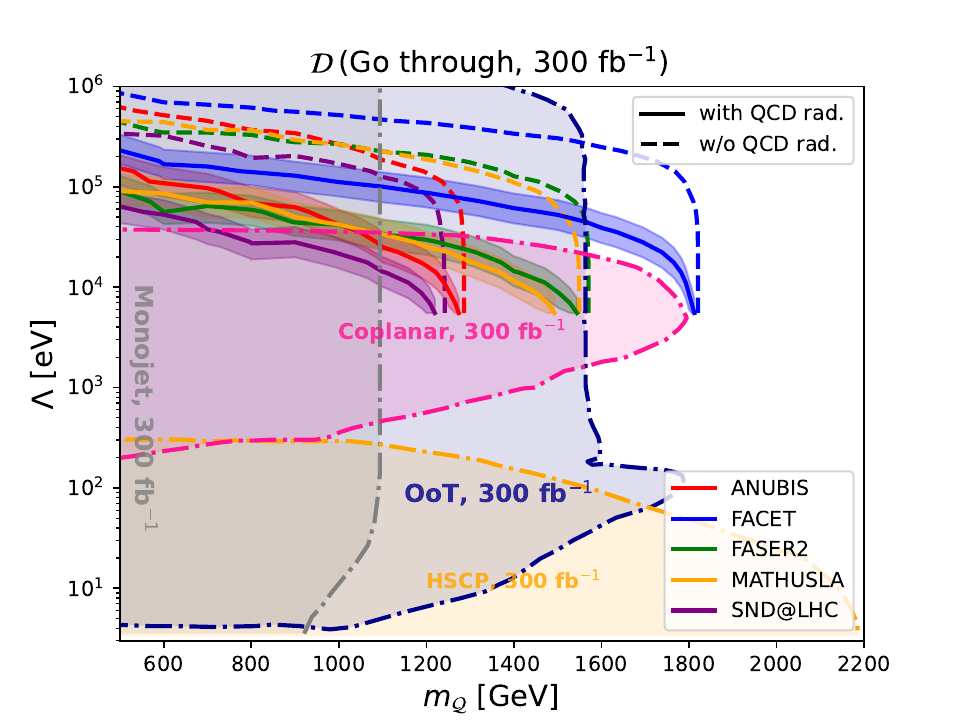}\\
\includegraphics[width=0.48\textwidth]{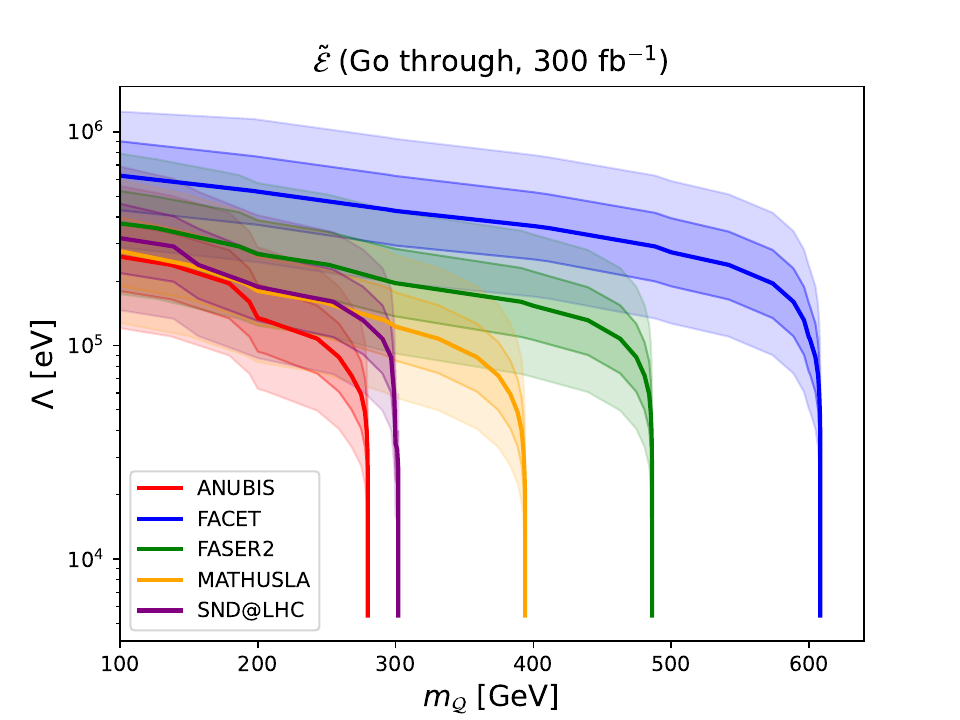}
\includegraphics[width=0.48\textwidth]{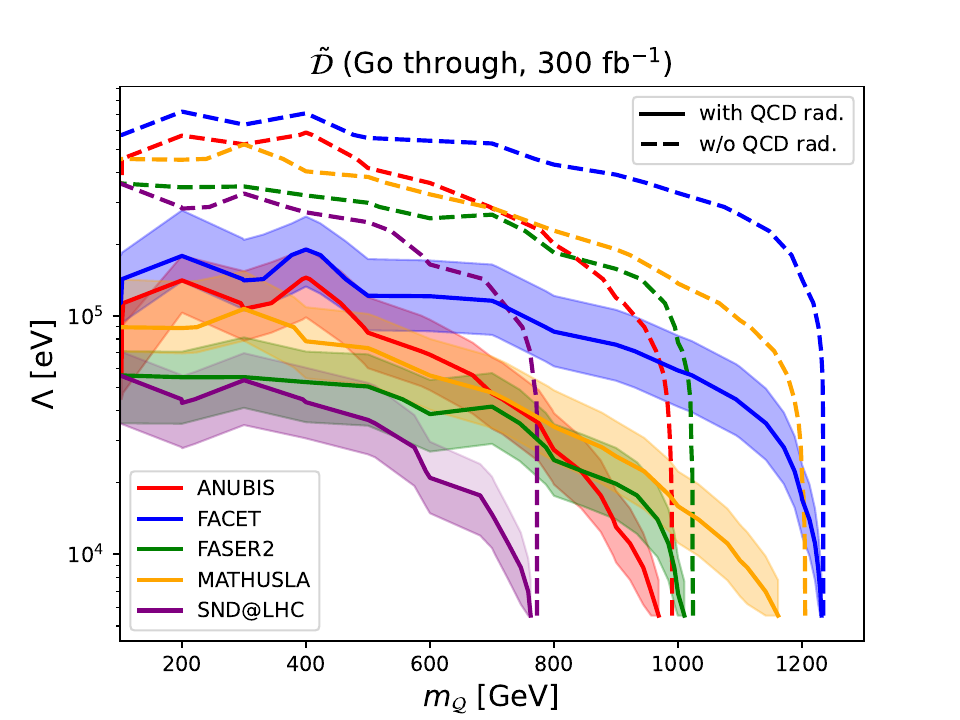}
\caption{{
Three-event contours for various quirk scenarios that have quirk going through different detectors (300 fb$^{-1}$ luminosity).
Uncolored Quirks ($\mathcal{E}$, $\tilde{\mathcal{E}}$): Solid curves represent contours based on $\epsilon=0.1$, with colors indicating detectors. Dark shaded areas show $\epsilon$ variability from $0.1/3$ to $3\times 0.1$ (larger $\epsilon$ implies smaller $\Lambda$), while light shaded areas indicate $\epsilon$ fluctuations from $0.1/10$ to $10 \times 0.1$.
Colored Quirks ($\mathcal{D}$, $\tilde{\mathcal{D}}$): Solid curves (with QCD radiation, $\epsilon^\prime=0.01$) and dashed curves (without QCD radiation, $\epsilon^\prime=0$) outline contours, with colors indicating detectors. Shaded areas around solid curves reflect $\epsilon^\prime$ variability from $0.01/2$ to $2\times 0.01$ (larger $\epsilon^\prime$ implies smaller $\Lambda$).
For two fermionic quirks ($\mathcal{E}$ and $\mathcal{D}$), the projected bounds from HSCP search~\cite{Farina:2017cts}, mono-jet search~\cite{Farina:2017cts}, coplanar search~\cite{Knapen:2017kly}, and out-of-time decay search~\cite{Evans:2018jmd} are shown by orange shaded region, gray shaded region, pink shaded region, and blue shaded region, respectively.}
\label{fig.bound}}
\end{figure}

\begin{figure}[thbp]
\includegraphics[width=0.48\textwidth]{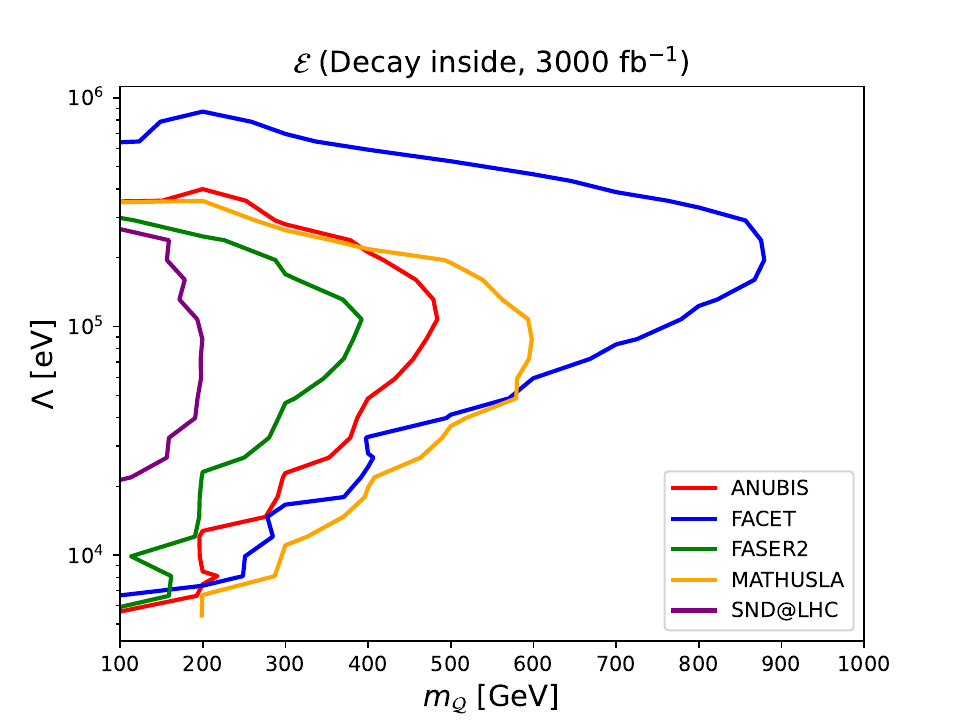}
\includegraphics[width=0.48\textwidth]{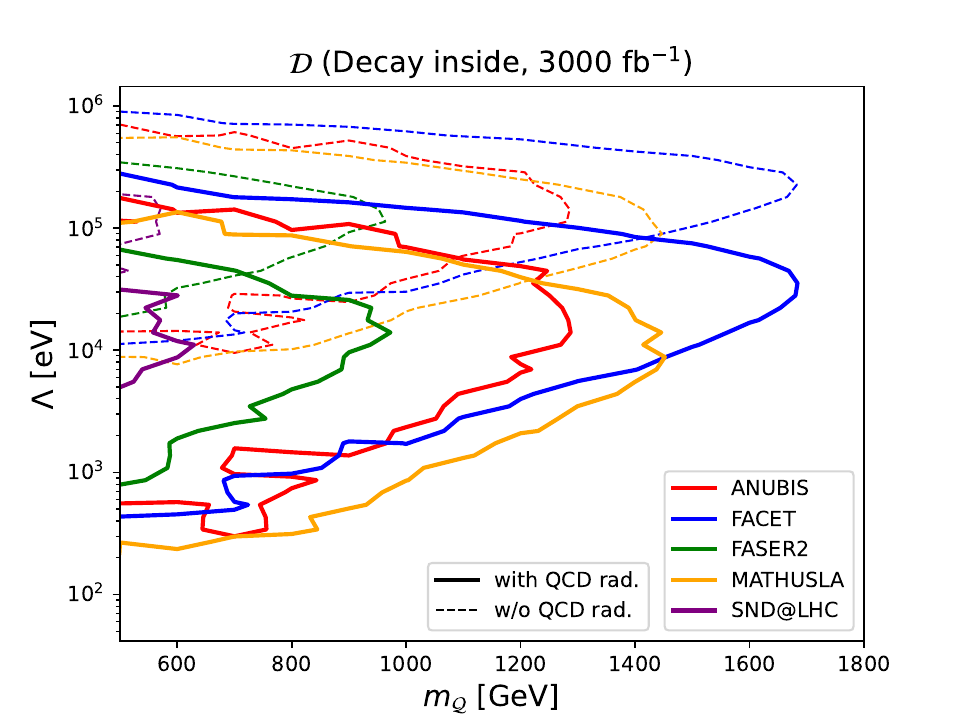}\\
\includegraphics[width=0.48\textwidth]{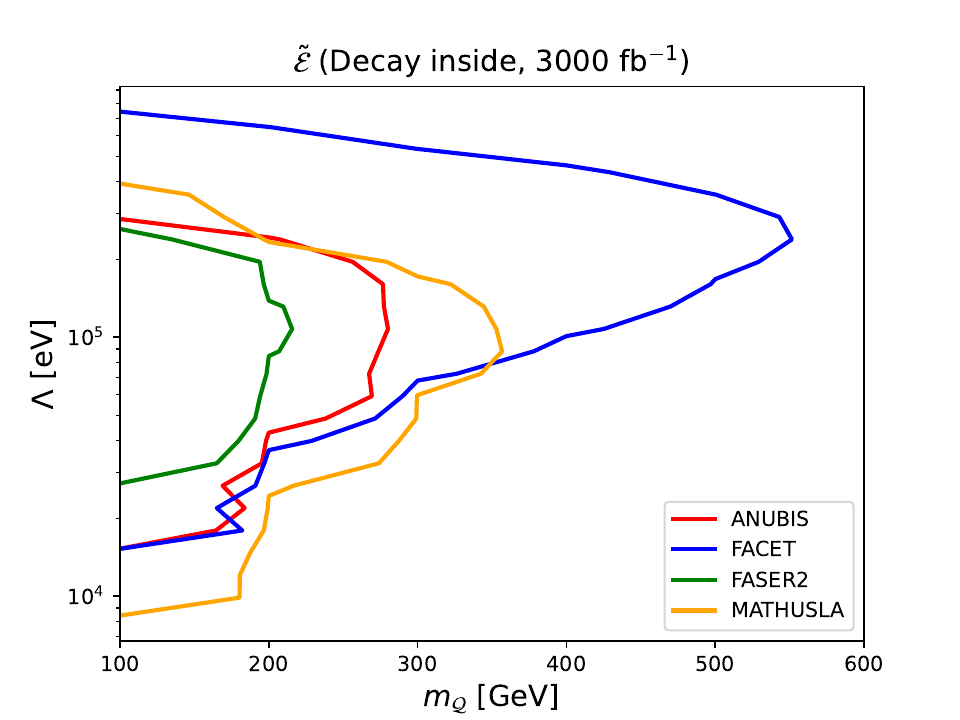}
\includegraphics[width=0.48\textwidth]{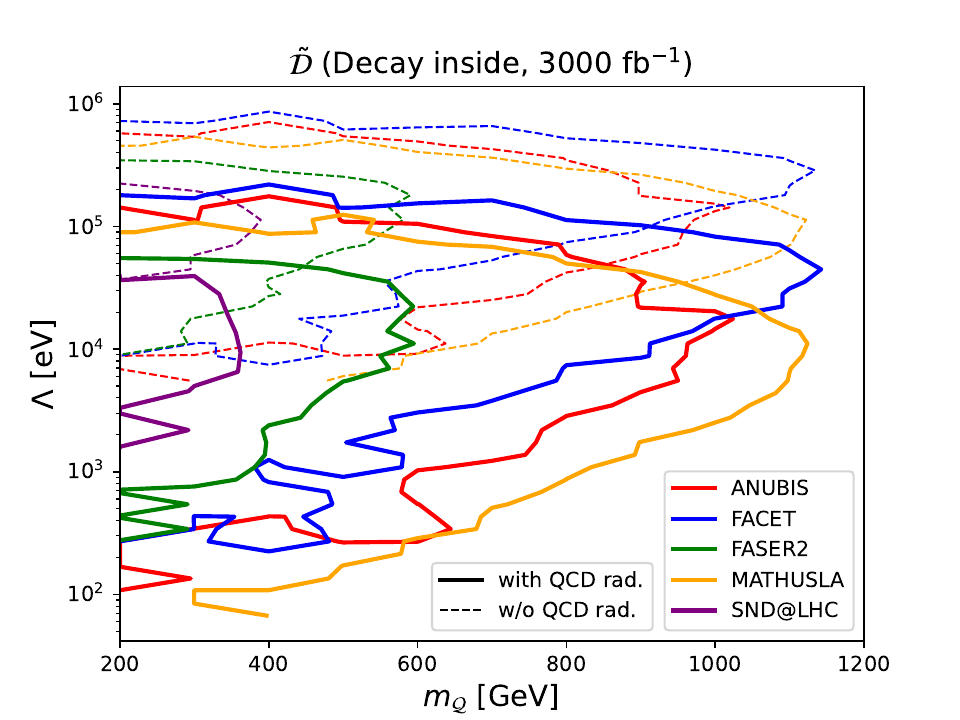}
\caption{{
Three-event contours for various quirk scenarios that have quirk decaying inside different detectors (3000 fb$^{-1}$ luminosity). Uncolored quirks ($\mathcal{E}$, $\tilde{\mathcal{E}}$): Solid curves represent contours based on $\epsilon=0.1$, with colors indicating detectors.
Colored quirks ($\mathcal{D}$, $\tilde{\mathcal{D}}$): Solid curves (with QCD radiation, $\epsilon^\prime=0.01$) and dashed curves (without QCD radiation, $\epsilon^\prime=0$) outline contours, with colors indicating detectors.}
\label{fig.boundin}}
\end{figure}

{In Fig.~\ref{fig.bound}, we present the three-event contours of the ``go through" quirk signature in the $\tilde{\mathcal{D}}$, $\tilde{\mathcal{E}}$, $\mathcal{D}$ and $\mathcal{E}$ quirk scenarios with an integrated luminosity of 300 fb$^{-1}$.
For uncolored quirks, $\mathcal{E}$ and $\tilde{\mathcal{E}}$, the solid curves delineating the three-event contours are set at an $\epsilon$ value of 0.1, with distinct colors denoting different detectors. The dark shaded region surrounding each curve highlights $\epsilon$ variability, from $0.1/3$ to $3\times 0.1$, illustrating that higher $\epsilon$ values are associated with lower $\Lambda$ values. The light shaded area adjacent to each curve indicates the fluctuation range of $\epsilon$, spanning from $0.1/10$ to $10 \times 0.1$.
Regarding colored quirks, $\mathcal{D}$ and $\tilde{\mathcal{D}}$, the solid curves incorporate QCD radiation effects by setting $\epsilon^\prime$ to 0.01, while the dashed curves omit these effects with $\epsilon^\prime$ at 0. Both types of curves, which are based on an $\epsilon$ value of 0.1, use varied colors to identify different detectors. The shaded regions around the solid curves reflect the variability in $\epsilon^\prime$, ranging from $0.01/2$ to $2\times 0.01$, with larger $\epsilon^\prime$ values indicating smaller $\Lambda$ values.
The shapes of the exclusion bounds of different detectors are similar. The FACET detector provides the most sensitive probe mainly due to its large effective opening angle to the IP. 
The detectors quickly lose the sensitivity when confinement scale $\Lambda \gtrsim 1$ MeV, because the energy loss is too efficient and the lifetime becomes too short.  
For $\Lambda \lesssim 100$ keV, the travel distance of the quirk reaches $\gtrsim \mathcal{O}(100)$ m which is around the detector distances to the IP. So, the exclusion limit becomes stable for the ``go through" signature in this parameter region. 
Comparing the sensitivities of MATHUSLA and FASER2, FASER2 has better sensitivity than MATHUSLA for the ``go through" signature, except for the $\tilde{D}$ scenario, in which the fraction of forward events is too low.  
The bounds on scalar quirks are much weaker than those on fermionic quirks when the gauge representations are the same mainly because of their smaller production rates.
As a result, we conclude that at LHC with integrated luminosity of 300 fb$^{-1}$, by using the ``go through" quirk signature,  the FACET detector will be able to probe the $\mathcal{E}$, $\mathcal{D}$, $\tilde{\mathcal{E}}$ and $\tilde{\mathcal{D}}$ quirks with mass up to 1000 GeV, 1800 GeV, 610 GeV, and 1240 GeV, respectively. 
The projected bounds from the Heavy Stable Charged Particle (HSCP) search, the mono-jet search, the coplanar search, and the out-of-time decay searches for fermionic quirks are shown as well. Those bounds are provided in literature with an integrated luminosity of 300 fb$^{-1}$.
It can be noted that the FACET and FASER2 detectors exhibit superior sensitivity compared to the majority of searches conducted at the LHC main detector, with the exception of the HSCP search in the $\mathcal{E}$ scenario. 
In the case of the colored quirk $\mathcal{D}$, only the FACET detector has the potential to offer a more effective probe than the array of searches performed at the LHC main detector, again excluding the HSCP search.
}

{
The configuration of the contours is influenced by the lifetime of the quirk-pair system $\tau_\text{tot}$, the dimensions of the detector, and the distance from the detector to the IP. When QCD radiation is disregarded and the value of $\epsilon$ is adjusted to $k\epsilon$, it becomes necessary to alter the value of $\Lambda$ to $k^{-1/3}\Lambda$ in order to maintain the same lifetime. This adjustment is required because $\tau^{\text{IC}}_{\text{linear}}$, as elaborated in Eq.~(\ref{eq:tlic}), significantly dictates the overall lifetime, $\tau_\text{tot}$. Consequently, the contours will shift along the Y-axis, as shown by the dark and light shaded regions surrounding each curve at scenarios of uncolored quirks, $\mathcal{E}$ and $\tilde{\mathcal{E}}$, in Fig.~\ref{fig.bound}. 

Considering QCD radiation, the emission of infracolor glueballs and QCD radiation within the linear quirk potential regime emerge as the key factors influencing $\tau_\text{tot}$. As delineated in Eqs.~(\ref{eq:tlic}), (\ref{eq:tlic2}), and (\ref{final}), $\tau$ is inversely proportional to $(\epsilon^\prime E_\text{QCD}\Lambda^2+\epsilon\Lambda^3)$.
Therefore, at scenarios of colored quirks, $\mathcal{D}$ and $\tilde{\mathcal{D}}$, with the inclusion of QCD radiation, the dashed curves will shift toward the solid ones with lower values of $\Lambda$, as shown in Fig.~\ref{fig.bound}. This shift is quantified by a translation coefficient of $\left(1+\frac{\epsilon^\prime E_\text{QCD}}{\epsilon \Lambda}\right)^{-1/3}$, where $\Lambda$ corresponds to the value at the solid curve. 
By setting $\epsilon=0.1$ and modifying $\epsilon^\prime=0.01$ by a factor of 2, the translation coefficient for the contours, corresponding to $\Lambda$ values ranging from $10^{-6}$ GeV to $10^{-3}$ GeV in the plots for colored quirks, will fluctuate between $\sqrt{2}$ and 1.32.
}

{In Fig.~\ref{fig.boundin}, we showcase the three-event contours for the ``decay inside" quirk signature across the scenarios of $\tilde{\mathcal{D}}$, $\tilde{\mathcal{E}}$, $\mathcal{D}$, and $\mathcal{E}$ quirks, assuming an integrated luminosity of 3000 fb$^{-1}$. For uncolored quirks, $\mathcal{E}$ and $\tilde{\mathcal{E}}$, the solid curves delineating the three-event contours are set against an $\epsilon$ value of 0.1, with different colors used to signify various detectors.
In the context of colored quirks, $\mathcal{D}$ and $\tilde{\mathcal{D}}$, the analysis includes both solid curves, which account for QCD radiation effects by setting $\epsilon^\prime$ to 0.01, and dashed curves, which omit QCD radiation with $\epsilon^\prime$ at 0. These curves, based on an $\epsilon$ value of 0.1, are distinguished by different colors to represent various detectors.
As previously noted, when the value of $\Lambda$ is less than or approximately 100 keV, the travel distance of the quirks can exceed or be on the order of 100 meters. This distance is roughly equivalent to the span from the interaction point (IP) to the detectors. It is within this parameter space that the ``decay inside" signature achieves its optimal detection potential.
Upon comparing the detection sensitivities of MATHUSLA and FASER2, it is evident that MATHUSLA consistently outperforms FASER2 in detecting the ``decay inside" signature, attributed to its significantly larger effective detector volume. Similarly, the constraints on scalar quirks are considerably more lenient than those on fermionic quirks when their gauge representations are identical.
In the scenarios involving colored quirks, $\mathcal{D}$ and $\tilde{\mathcal{D}}$, the transition from dashed to solid curves—indicative of including QCD radiation effects—can be elucidated in a manner akin to the explanations provided for Fig.~\ref{fig.bound}.
Conclusively, at the HL-LHC with an integrated luminosity of 3000 fb$^{-1}$, utilizing the ``decay inside" quirk signature enables the FACET detector to probe quirks in the $\mathcal{E}$, $\mathcal{D}$, $\tilde{\mathcal{E}}$, and $\tilde{\mathcal{D}}$ scenarios with masses up to 880 GeV, 1680 GeV, 550 GeV, and 1140 GeV, respectively. It should be noted that adjusting the emission probabilities, $\epsilon$ and $\epsilon^\prime$, results in an overall upward or downward shift of the bounds by the same amount as demonstrated in the ``go through" signature cases depicted in Fig.~\ref{fig.bound}.}


\begin{figure}[htbp]
\includegraphics[width=0.48\textwidth]{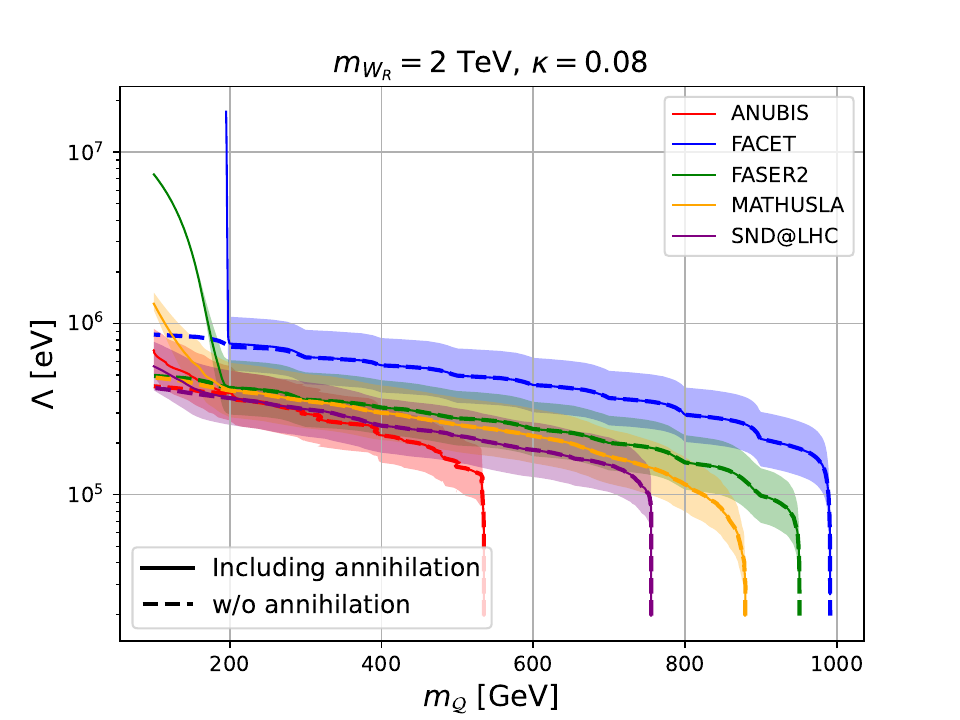}
\includegraphics[width=0.48\textwidth]{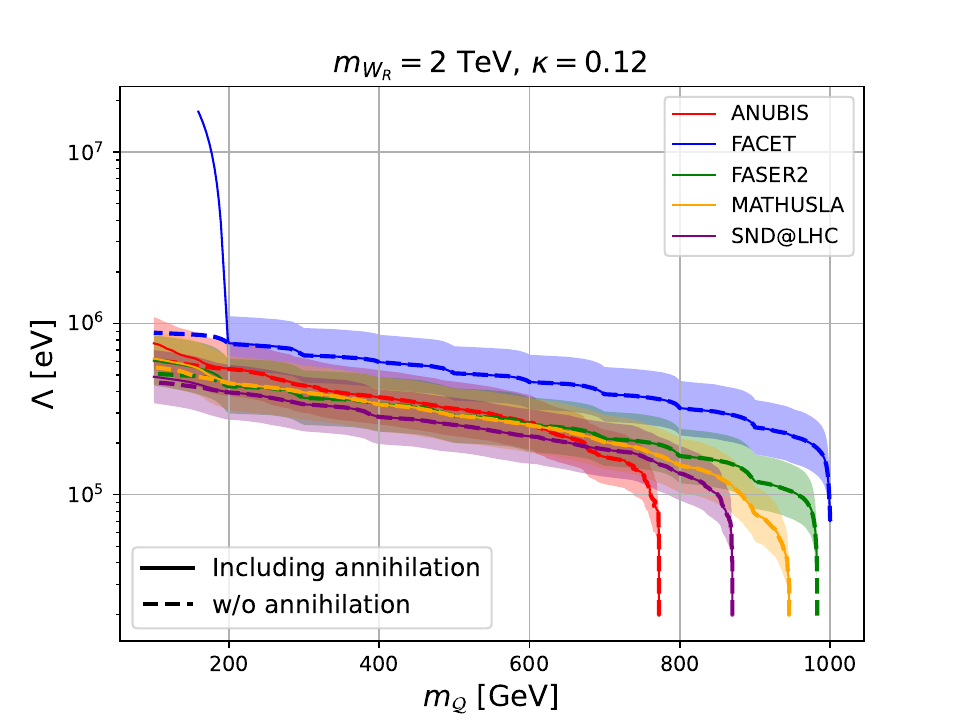}\\
\includegraphics[width=0.48\textwidth]{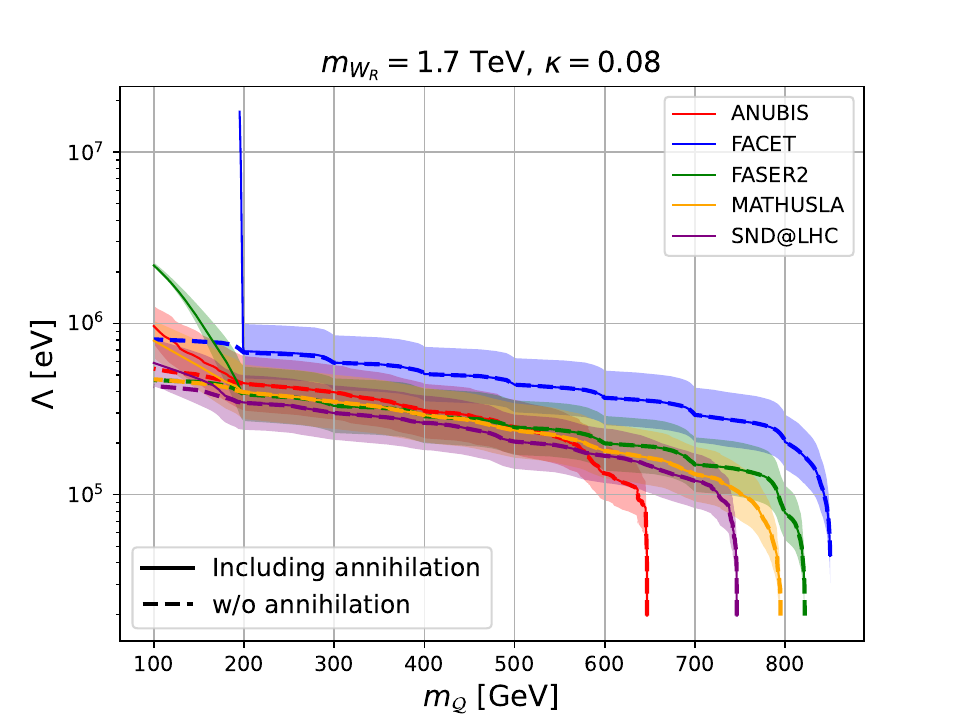}
\includegraphics[width=0.48\textwidth]{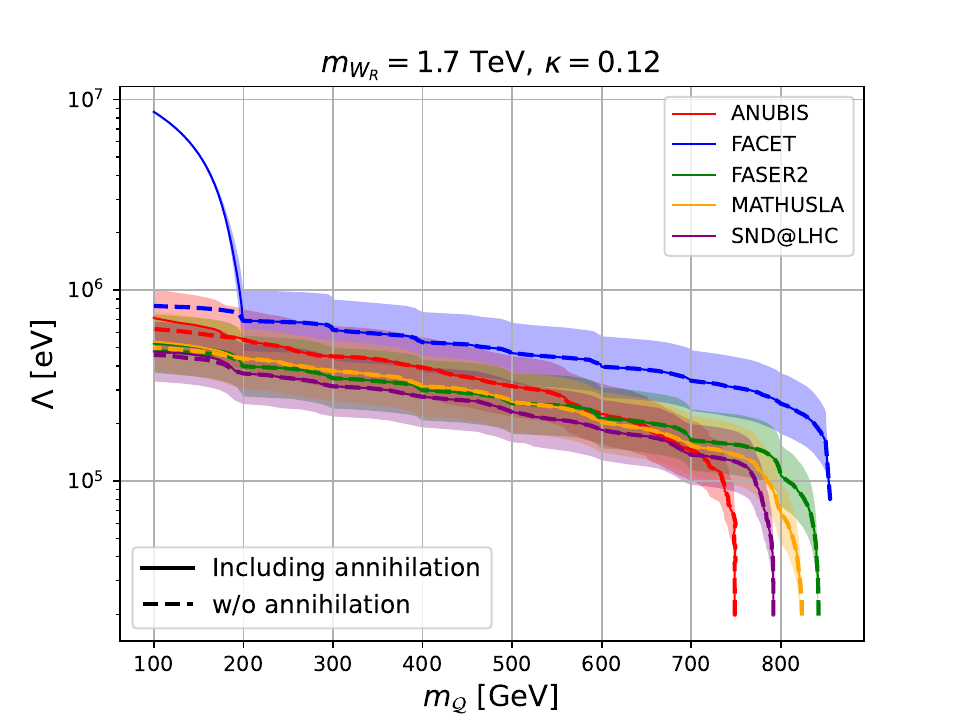}
\caption{{Three-event contours of the $W_R$ quirk scenarios that have quirk going through different detectors. Solid curves show results including quirk pair annihilation time, while dashed curves exclude it, both at $\epsilon=0.1$. Shaded areas around solid curves reflect $\epsilon$ variability from $0.1/3$ to $3\times 0.1$ (larger $\epsilon$ implies smaller $\Lambda$). The $W_R$ mass and $\kappa$ are given in the title of each plot.} 
\label{fig.boundwr}}
\end{figure}

{The three-event contours for the $W_R$ quirk scenario with different parameter setups are presented in Fig.~\ref{fig.boundwr}. 
The solid curves depict results that account for the quirk pair annihilation time, whereas the dashed curves represent scenarios where this factor is omitted, with both sets of curves based on an $\epsilon$ value of 0.1. The shaded regions surrounding the solid curves illustrate the variability in $\epsilon$, ranging from $0.1/3$ to $3\times 0.1$, indicating that a larger $\epsilon$ corresponds to a smaller $\Lambda$.
The annihilation effects are important only for $\Lambda \gtrsim 1$ MeV when its time scale is comparable to or larger than the period of radiative energy loss. The region with a confinement scale approaching the GeV scale may still be probed. 
The annihilation lifetime drops dramatically with increasing the quirk mass and become irrelevant for $m_{\mathcal{Q}} \gtrsim 200$ GeV. 
The sensitivities of those detectors are ranked similarly to color neutral $\mathcal{E}$ and $\tilde{\mathcal{E}}$ scenarios. 
Heavier $W_R$ and smaller $\kappa$ lead to a longer-lived quirk pair as given in Eq.~(\ref{eq:tani}), thus having better probing prospects. 
In the small $\Lambda$ region, where the period of radiative energy loss is dominating over the annihilation lifetime, the changes of $m_{W_R}$ and $\kappa$ lead to different production kinematics and production rates, thus giving different reach limits. 
In this scenario, both the FASER2 and FACET will be able to probe the quirk with mass up to around half of the $W_R$ mass.} 

\section{Discussion and conclusion} \label{sec:conclude}

This work considers the probing prospect of the LHC far detectors to five quirk scenarios. 
Each of the $\tilde{\mathcal{D}}$, $\tilde{\mathcal{E}}$, $\mathcal{D}$ and $\mathcal{E}$ scenarios contains a single quirk with different quantum numbers. The $W_R$ scenario contains two scalar quirks and a new gauge boson $W_R$. 
The NLO QCD corrections are included in our simulation. Those effects can not only increase the production rate by a factor of $\sim (1.2-1.4)$ for all scenarios but also significantly change the quirk kinematics. In particular, the fraction of quirk events in the forward direction is enhanced. 

After being pair-produced at the LHC, the quirk loses its kinetic energy through the radiations of infra-color glueballs and photons, and settles into a ground state which subsequently decays through the constituent quirk annihilation. 
In the $\tilde{\mathcal{D}}$, $\tilde{\mathcal{E}}$, $\mathcal{D}$ and $\mathcal{E}$ scenarios, the annihilation of the quirk pair into infracolor glueballs is unavoidable such that the lifetime of annihilation is always orders of magnitude smaller than the period of radiative energy loss for $\Lambda \lesssim \mathcal{O}(1)$ MeV.
The $W_R$ scenario is proposed to illustrate the interface between quirk annihilation and radiative energy loss. 
For the production of $\tilde{\mathcal{E}} \tilde{\mathcal{V}}$ quirk pair in $W_R$ scenario, the annihilation into infracolor glueball is forbidden by the flavor symmetry. Moreover, the direct annihilations into SM fermions are $p$-wave suppressed. So, the quirk pair can only annihilate into $W_R^{(*)} \gamma$, the cross-section of which is suppressed by $(\kappa/m_{W_R})^4$. As a result, the annihilation lifetime can be comparable to or even longer than the period of radiative energy loss in some parameter space. 
The $\beta$ decay of the $\tilde{\mathcal{E}} \tilde{\mathcal{V}}$ bound state is ignored (which is true as long as the mass difference is smaller than $\sim 0.3$ GeV) since it tends to reduce to the annihilation lifetime. 

The LHC far detectors considered in this work include FASER2, SND@LHC, ANUBIS, FACET, and MATHUSLA. 
Both quirk signatures with the quirk pair decaying inside the effective detector volume and passing through the detector are studied. 
Among those far detectors, FACET provides the best sensitivity as it has the largest opening angle to the IP. However, the backgrounds of the FACET detector are much heavier than those of other detectors. 
In such a case, 3 signal events may not be enough to claim an exclusion. The usage of specific features may be essential for suppressing the background and identifying the quirk signal. 
For the ``go through" quirk signature, the FASER2 has better sensitivity than the MATHUSLA for all scenarios except the $\tilde{\mathcal{D}}$ scenario, in which the fraction of events with forward propagating quirk pair is highly suppressed. 
While for the ``decay inside" quirk signature, MATHUSLA exhibits better sensitivity, because of its larger effective detector volume. 
If the background can indeed be controlled to a negligible level, the ``go through" signature always provides a better probe, although the ``decay inside" signature is more remarkable and easier for identification. 
With integrated luminosity of 300 fb$^{-1}$, the search for ``go through" signature at the FACET detector will be able to probe the $\mathcal{E}$, $\mathcal{D}$, $\tilde{\mathcal{E}}$ and $\tilde{\mathcal{D}}$ quirks with {mass up to  1000 GeV, 1800 GeV, 610 GeV, and 1240 GeV}, respectively.


In the $W_R$ scenario, the annihilation lifetime is important to the signal efficiency only in the region with $\Lambda \gtrsim 0.4$ MeV and $m_{\mathcal{Q}} \lesssim 200$ GeV, because the annihilation lifetime decreases dramatically with increasing quirk mass and the period of energy loss is dominating over the annihilation time in the small $\Lambda$ region. 
In the cases with a dominating annihilation lifetime, the region with a confinement scale as high as GeV can be probed by the FACET detector. The reach limit to the $\Lambda$ is higher for heavier $W_R$ and smaller $\kappa$. In the small $\Lambda$ region, both the FASER2 and FACET will be able to probe the quirk with mass up to around half of the $W_R$ mass.


\appendix
\section{Derivation For $d_\text{min}$} \label{dp}
{
We denote the magnetic field in the Lab frame as $\vec{B}$. After a single oscillation of the quirk pair in $\vec{B}$, the angular momentum of the quirk-pair system within the CoM frame can be induced by the electric field present in the CoM frame (resulting from the Lorentz transformation of $\vec{B}$ from the Lab frame to the CoM frame), denoted as 
  \begin{align}
      \vec{L}^\prime=2q E^\prime\int_{0}^{t^\prime_p}  r(t^\prime)dt^\prime=p^\prime d_\text{min}^\prime~.\label{LL}
  \end{align}
Here, $p^\prime$, $t^\prime$, and $d_\text{min}^\prime$ correspond to the initial momentum magnitude, time, and the minimum separation distance between the two quirks within the CoM frame, respectively. The term $r(t^\prime)$ specifies the position of the quirk at time $t^\prime$, while $t^\prime_p$ represents the duration of one oscillation within the CoM frame. $E^\prime$ signifies the component of the electric field oriented perpendicular to the motion of the quirk within the CoM frame. As stated in \cite{Li:2020aoq}, it follows that
\begin{align}
    r(t^\prime)&=\frac{m_\mathcal{Q}}{\Lambda^2}\left(\sqrt{1+\rho^2}-\sqrt{1+\rho^2\left(1-\frac{\Lambda^2}{m_\mathcal{Q}\rho}t^\prime\right)^2}\right)~,\label{rr}\\
    t^\prime_p&=\frac{2m_\mathcal{Q}\rho}{\Lambda^2}~,\label{tt}
\end{align}
where $\rho$ is defined in Sec.~\ref{ppp}.
By applying a Lorentz transformation to the magnetic field $\vec{B}$ from the Lab frame to the CoM frame, we obtain
\begin{align}
    E^\prime &=\frac{\left|q\left(\vec{\beta}\times \vec{B}\right)\cdot\hat{e}_3\right|}{\sqrt{1-\beta^2}}~.\label{ee}
\end{align}
The variables present in Eq.~(\ref{ee}) are delineated in Section~\ref{ppp}.
By substituting Eqs.~(\ref{rr}), (\ref{tt}), and (\ref{ee}) in Eq.~(\ref{LL}), we get 
\begin{align}
    d^\prime_{\text{min}}=& 0.231\times 10^{11}[\mu\text{m}]\nonumber\\
    & \times{\frac{m_\mathcal{Q}}{[\text{GeV}]} \left(\frac{\Lambda}{[\text{eV}]}\right)^{-4}\left(\sqrt{1+\rho^2}-\frac{\sinh^{-1}{\rho}}{\rho}\right)} {\frac{\left|q\left(\vec{\beta}\times \frac{\vec{B}}{[\text{T}]}\right)\cdot\hat{e}_3\right|}{\sqrt{1-\beta^2}}}~.\label{dminp}
\end{align}
The minimum distance $d^\prime_{\text{min}}$ in the CoM frame can be decomposed into components that are parallel and perpendicular to the direction of the quirk-pair motion (the direction of the Lorentz transformation), denoted as 
  \begin{align}
     &d^\prime_\parallel=d_\text{min}^\prime\cos\alpha~,\\
     &d^\prime_\perp=d_\text{min}^\prime\sin\alpha~,
  \end{align}
  where $\cos\alpha$ is given in Eq.~(\ref{alpha}).
 Following the Lorentz transformation from the CoM frame to the Lab frame, we obtain
 \begin{align}
     &d_\parallel=d^\prime_\parallel/\sqrt{1-\beta^2}~,\\
     &d_\perp=d^\prime_\perp~, \\
     & d_\text{min}=\sqrt{d_\parallel^2+d_\perp^2}=d_\text{min}^\prime\sqrt{1+\frac{\beta^2}{1-\beta^2}\cos^2\alpha}~.\label{a10}
  \end{align}
The minimum distance between two quirks following a single oscillation in the magnetic field can be determined through the labor-intensive simulation of quirk motions, following the methodology proposed in \cite{Li:2019wce}. This distance, ascertained via simulation, is denoted as $d^s_\text{min}$. Our findings reveal that by integrating an additional coefficient of 1.2 into Eq.~(\ref{a10}), we observe that approximately $88\%$ of quirk-pair events exhibit a ratio of $d_\text{min}/d^s_\text{min}$ falling within the range of [0.5, 1.5]. This refinement significantly enhances the accuracy of our distance estimations in the study of quirk-pair dynamics, as reflected in Eq.~\ref{dmin}.
}

\begin{acknowledgments}
We are grateful to Jonathan Feng for providing valuable comments on the results. 
This work was supported by the Natural Science Foundation of Sichuan Province under grant No. 2023NSFSC1329 and the National Natural Science Foundation of China under grant No. 11905149 and No.12247119.

\end{acknowledgments}

\bibliography{LLPquirk}

\end{document}